\documentclass[a4paper,notitlepage,eqsecnum,nofootinbib]{revtex4-1}
\usepackage{slashed,bbold,bm,amsmath,amssymb,mathtools,wrapfig,epsfig} 
\usepackage[usenames]{color}
\usepackage{hyperref}
\usepackage{tcolorbox}
\hypersetup{
    colorlinks=true,       
    linkcolor=red,          
    citecolor=blue,        
    filecolor=green,      
    urlcolor=magenta           
}

\newcommand{\crt}{\\[2mm]}
\newcommand{\nn}{\nonumber}
\newcommand{\beq} {\begin{equation}}
\newcommand{\eeq} {\end{equation}}
\newcommand{\beqa} {\begin{eqnarray}}
\newcommand{\eeqa} {\end{eqnarray}}

\newcommand{\bs}[1]{\boldsymbol{#1}}

\newcommand{\ie}{{\it i.e.}}

\newcommand{\eg}{{\it e.g.}}

\newcommand{\as}{\alpha_s}

\newcommand{\la}{\Lambda}

\newcommand{\veps}{\varepsilon}
\newcommand{\order}[1]{${\cal O}\left(#1 \right)$}
\newcommand{\morder}[1]{{\cal O}\left(#1 \right)}
\newcommand{\eq}[1]{(\ref{#1})}
\newcommand{\fig}[1]{Fig.~\ref{#1}}
\newcommand{\gsim}{\gtrsim}   

\newcommand{\inv}[1]{\frac{1}{#1}}
\newcommand{\halft}{{\textstyle \frac{1}{2}}}
\newcommand{\quart}{{\textstyle \frac{1}{4}}}

\newcommand{\intt}{{\textstyle \int}}
\newcommand{\ket}[1]{\left\vert{#1}\right\rangle}

\newcommand{\bra}[1]{\langle{#1}\vert}

\newcommand{\com}[2]{\left[{#1},{#2}\right]}
\newcommand{\comb}[2]{\big[{#1},{#2}\big]}
\newcommand{\acom}[2]{\left\{{#1},{#2}\right\}}
\newcommand{\acomb}[2]{\big\{{#1},{#2}\big\}}
\newcommand{\tr}{\mathrm{Tr}\,}

\newcommand{\mC}{\mathcal{C}}

\newcommand{\mE}{\mathcal{E}}

\newcommand{\mG}{\mathcal{G}}
\newcommand{\mH}{\mathcal{H}}

\newcommand{\mL}{\mathcal{L}}

\newcommand{\mO}{\mathcal{O}}
\newcommand{\mP}{\mathcal{P}}
\newcommand{\bP}{\mathbb{P}}
\newcommand{\mS}{\mathcal{S}}

\newcommand{\xv}{{\bs{x}}}

\newcommand{\Ev}{{\bs{E}}}

\newcommand{\gz}{\gamma^0}
\newcommand{\go}{\gamma^1}

\newcommand{\qb}{{\bar{q}}}
\newcommand{\pb}{{\bar{p}}}
\newcommand{\eb}{{\bar{e}}}
\newcommand{\sbb}{{\bar{s}}}

\newcommand{\kum}{{\,{_1}F_1}}

\newcommand{\xb}{{x_{B}}}
\newcommand{\yb}{{y_{B}}}

\newcommand{\rar}{\rightarrow}
\newcommand{\lar}{\leftarrow}
\newcommand{\lra}{\leftrightarrow}
\newcommand{\rder}{{\buildrel\rar\over{\partial}}}
\newcommand{\lder}{{\buildrel\lar\over{\partial}}}
\newcommand{\lrder}{{\buildrel\lra\over{\partial}}}

\newcommand{\phip}{\phi^{\scriptscriptstyle{(P)}}}
\newcommand{\phir}{\phi^{\scriptscriptstyle{(0)}}}
\newcommand{\phii}{\phi^{\scriptscriptstyle{(\infty)}}}

\newcommand{\Phir}{\Phi^{\scriptscriptstyle{(0)}}}
\newcommand{\Phip}{\Phi^{\scriptscriptstyle{(P)}}}
\newcommand{\Phii}{\Phi^{\scriptscriptstyle{(\infty)}}}
\newcommand{\Ui}{U^{\scriptscriptstyle{(\infty)}}}
\newcommand{\Vi}{V^{\scriptscriptstyle{(\infty)}}}
\newcommand{\Ti}{\Theta^{\scriptscriptstyle{(\infty)}}}
\newcommand{\taui}{\tau_{\scriptscriptstyle{\infty}}}

\newcommand{\dx}{\partial_x}

\newcommand{\dt}{\partial_t}

\def\Xint#1{\mathchoice
   {\XXint\displaystyle\textstyle{#1}}%
   {\XXint\textstyle\scriptstyle{#1}}%
   {\XXint\scriptstyle\scriptscriptstyle{#1}}%
   {\XXint\scriptscriptstyle\scriptscriptstyle{#1}}%
   \!\int}
\def\XXint#1#2#3{{\setbox0=\hbox{$#1{#2#3}{\int}$}
     \vcenter{\hbox{$#2#3$}}\kern-.5\wd0}}

\def\dashint{\Xint-}

\begin{document}

\title{'t Hooft model in the temporal gauge}

\author{Paul Hoyer}
\affiliation{ \vspace{1mm} Department of Physics, POB 64, FIN-00014 University of Helsinki, Finland}
\email{paul.hoyer@helsinki.fi}

\begin{abstract}  
I consider QCD$_2$ in the $N_c \to \infty$ limit at fixed $g^2N_c$. The derivation starts from equal-time $q\qb$ bound states in coordinate space and temporal ($A^0=0$) gauge, avoiding the use of quark and gluon propagators. The wave function is given analytically by a $_1F_1$ function with an explicit frame dependence. In the infinite momentum frame the Fourier transformed wave function satisfies the 't~Hooft equation, however with contributions also from quarks with negative kinetic energy. 
Such contributions are present also in the rest frame, and do not vanish under boosts.
\end{abstract}

\maketitle

\vspace{-.5cm}

\parindent 0cm

\section{Introduction} \label{sec1} 

The solution of QCD in $D=1+1$ dimensions (QCD$_2$) found by 't Hooft in 1974 \cite{tHooft:1974pnl} has inspired many studies and allowed insights into hadron and confinement dynamics (see \cite{Callan:1975ps, *Einhorn:1976uz,*Einhorn:1976ax,*Brower:1977hx} for some early contributions). 't Hooft used light front quantization (states defined at equal $x^+ =t+x$) in $A^+=A^0+A^1=0$ gauge, taking the limit of a large number of colors, $N_c \to \infty$ at fixed $g^2N_c$. Planar Feynman diagrams were found to dominate, and bound states to be stable at leading order of $1/N_c$ \cite{tHooft:1973alw,*Witten:1979kh}.

In 1978 Bars and Green \cite{Bars:1977ud} managed to derive 't Hooft's solution by quantizing QCD$_2$ at equal ordinary time $t$ in Coulomb (Axial) gauge, $A^1=0$. Special care was required since the confined quark propagator is singular and not Poincar\'e covariant. 't~Hooft's result for the bound state wave functions was reproduced in the infinite momentum frame (IMF). Poincar\'e generators were found which satisfy the Lie algebra when operating on color singlet bound states. 
The boost covariance of the solution \cite{Bars:1977ud} was confirmed numerically, first in the rest frame \cite{Li:1987hx} and more recently in a general frame \cite{Jia:2017uul}. There are also lattice studies of QCD$_2$ \cite{Hamer:1981yq,*Berruto:2002gn,*GarciaPerez:2016hph}.

The QCD$_2$ action is
\begin{align} \label{1.1}
\mS_{QCD} &= \int dt\,dx\big[\halft F_a^{01}F_a^{01} + \bar\psi(i\slashed{\partial}-m-g A_a^0\gamma^0 T^a+g A_a^1\gamma^1 T^a)\psi\big] \hspace{2cm} F^{01}_a = \dt A_a^1+\dx A_a^0 - gf_{abc}A_b^0 A_c^1 \nn\crt
&= \int dt\,dx\big[\halft (\dt A_a^1)^2+\bar\psi(i\slashed{\partial}-m+g A_a^1\gamma^1 T^a)\psi\big] \hspace{2cm} (\mbox{temporal\ gauge,}\ A^0 \to 0)
\end{align}
The present study of the 't Hooft model uses quantization at equal time $t$ as in \cite{Bars:1977ud,Li:1987hx,Jia:2017uul}, but differs from previous work in other respects. 

\textit{(i) Temporal gauge, $A^0=0$}  \cite{Willemsen:1977fr,*Bjorken:1979hv,*Leibbrandt:1987qv,*Strocchi:2013awa}, which is advantageous for canonical quantization. Since both $A^0$ and its conjugate field vanish ($\mS_{QCD}$ is independent of $\dt A^0$) QCD can be canonically quantized without the Dirac constraints required in Coulomb gauge (see \cite{Christ:1980ku} and Sec. 8.2 of \cite{Weinberg:1995mt}). The color electric field is conjugate to $A_a \equiv A_a^1$ and $i\psi^\dag$ is conjugate to $\psi$,
\begin{align} \label{1.2}
E_a(t,x) = -\frac{\delta \mS_{QCD}}{\delta[\dt A_a(t,x)]} = -\dt A_a(t,x) \hspace{2cm}
i\psi^\dag(t,x) = \frac{\delta \mS_{QCD}}{\delta[\dt \psi(t,x)]}
\end{align}
giving the equal-time commutation relations,
\begin{align} \label{1.3}
\com{E_a(t,x)}{A_b(t,y)} &= i\delta_{a,b}\delta(x-y)  & \acomb{\psi^\dag_{A,\alpha}(t,x)}{\psi_{B,\beta}(t,y)} = \delta_{AB}\,\delta_{\alpha\beta}\,\delta(x-y)
\end{align}
The generators of space and time translations as well as boosts are (unlike in Coulomb gauge) local functions of the quark and gluon fields.
The $A^0=0$ condition fixes temporal gauge only up to time independent gauge transformations. Such transformations are generated by Gauss' operator 
\begin{align} \label{1.4}
\mG_a(t,x)  \equiv \frac{\delta\mS_{QCD}}{\delta{A_a^0(t,x)}} = \dx E_a(t,x)+g f_{abc}A_b E_c(t,x)-g\psi^\dag T^a\psi(t,x)
\end{align}
which commutes with the Hamiltonian \cite{Willemsen:1977fr,*Bjorken:1979hv,*Leibbrandt:1987qv,*Strocchi:2013awa}.
Gauss' law $\mG_a(t,x)=0$ is not an operator equation of motion in temporal gauge. Invariance under time independent gauge transformations is ensured by a \textit{constraint on physical states},
\begin{align} \label{1.5}
\mG_a(t,x)\ket{phys,t} =0  \hspace{1cm} \mbox{including} \hspace{1cm} \mG_a(t,x)\ket{0} =0
\end{align} 
This determines the divergence of the electric field $\dx E_a(t,x)$ in \eq{1.4} in terms of the gluons and quarks in $\ket{phys,t}$, as probed by their commutators with $g f_{abc}A_b E_c(t,x)-g\psi^\dag T^a\psi(t,x)$. Gauss' constraint \eq{1.5} on $\ket{phys,t}$ involves no particle creation from the vacuum.

\textit{(ii) Color singlets.} Previous studies of the 't Hooft model used (colored) quark and gluon propagators to build (color singlet) bound states via the Bethe-Salpeter equation \cite{Salpeter:1951sz}. Due to the confining interaction in $D=1+1$ dimensions the propagators need to be regularized, with the singularities canceling for the bound states. 

I avoid the propagators by starting from color singlet $q\qb$ states, defined at $t=0$ in coordinate space,
\begin{align}
\ket{M,P}&= \int dx_q dx_\qb\,\bar\psi(x_q)\mathbb{P}(x_\qb \to x_q)\,\exp\big[\halft iP(x_q+x_\qb)\big]\Phip(x_q-x_\qb)\psi(x_\qb)\ket{0} \hspace{1cm} (t=0)  \label{1.6} \crt
&\mathbb{P}(x_\qb \to x_q) \equiv \mathbb{P}\exp\Big[ig\int_{x_\qb}^{x_q}dx\,A_a(x)T^a\Big]
 \hspace{2cm} (\mathbb{P}:\ \mbox{path ordering}) \label{1.7}
\end{align}
where $M$ denotes the mass and $P \equiv P^1$ the momentum of the bound state. The gauge link $\mathbb{P}(x_\qb \to x_q)$ is required to satisfy \eq{1.5} (see below). The c-numbered wave function $\Phip(x_q-x_\qb)$ is a unit matrix in color space and a $2\times 2$ matrix in the Dirac indices. Nominally, $\bar\psi(x_q)$ creates a quark at $x_q$ and $\psi(x_\qb)$ an antiquark at $x_\qb$, both at $t=0$. In the absence of quark and gluon propagators there is no concept of ``planar diagrams''. 

Fixing the gauge over all space at the same time ($A_a^0(t,x)=0$) gives rise to the instantaneous potential derived below. The state \eq{1.6} is an eigenstate of the Hamiltonian if the wave function $\Phip$ satisfies the bound state equation (BSE) \eq{2.13}. This is an ordinary differential equation with an analytic solution. After a Fourier transform the BSE becomes an integral equation in momentum space, which (for $P \to \infty$) I compare with the equation found by 't Hooft.

\vspace{.3cm}

The rest of this paper is organized as follows. In Sec. \ref{sec2} and Appendixes \ref{appA} and \ref{appB} I verify that the states \eq{1.6} satisfy the constraint \eq{1.5} and determine the Poincar\'e generators of QCD$_2$ in temporal gauge. The states \eq{1.6} are demonstrated to be be eigenstates of the generator of space translations $\hat P^1$ with eigenvalue $P$, and of the Hamiltonian $H=\hat P^0$ when the BSE \eq{2.13} for the wave function is satisfied. The gauge link \eq{1.7} generates the linear potential,
\begin{align} \label{1.7a}
V(x) = V'|x| \hspace{2cm} V' = \halft g^2C_F = g^2\,\frac{N_c^2-1}{4N_c}
\end{align}
Section \ref{sec3} determines the properties of the wave function, including its symmetries under parity and charge conjugation.  $\Phip(x)$ is expressed in terms of the confluent hypergeometric function $_1F_1$ with argument $\tau_P$, the square of the ``kinetic momentum'' $(E_P-V,P)$,
\begin{align} \label{1.7b}
V'\tau_P(x) \equiv (E_P-V)^2-P^2 = M^2-2E_P V'|x| + V'^2x^2 \hspace{2cm} E_P = \sqrt{P^2+M^2}
\end{align}
Viewed as a function of $\tau_P$, $\Phip(\tau_P)$ is independent of $P$ up to a Dirac matrix. In Appendix \ref{appC} I verify that the frame dependence of $\Phip(\tau_P(x))$, viewed as a function of $x$, is consistent with that given by the boost generator $\hat M^{01}$.

In Sec. \ref{sec4} I consider the Fourier transform \eq{4.1} from coordinate to momentum space, $\Phip(x) \to \Phip(k)$. The wave function is then projected \eq{4.5} on components $\Phip_{s\sbb}$ according to the sign of the of the quark ($s$) and antiquark ($\sbb$) kinetic energy. The BSE in momentum space \eq{4.15} shows how the potential mixes the wave function components with positive and negative kinetic energies.

Sec. \ref{sec5} shows that the Fourier transform of $\Phip(x)$ can be done analytically in the IMF, $P \to \infty$. The IMF BSE \eq{5.15} has the form of the 't Hooft bound state equation \eq{5.16} \cite{tHooft:1974pnl}, but with contributions also from negative kinetic energies (at quark momentum fractions $\xb<0$ and $\xb>1$). An example IMF wave function is shown in \fig{f1} (right) and is numerically verified to solve the BSE \eq{5.15} in Table \ref{t1}. 

Returning to $x$-space in Sec. \ref{sec6} I note that the electric field $E_a$ of the $q\qb$ Fock state may be determined from Gauss' constraint \eq{1.5}, viewed as a differential equation for $\dx E_a$ \eq{1.4}. This \eq{6.5} gives the same linear potential \eq{1.7a} as previously. Solving the differential equation for $E_a$ allows (for color singlet states in QCD) to include a homogeneous solution which induces an isotropic gluon energy density $E_\la$ \eq{6.16}, akin to the ``bag constant'' \cite{Chodos:1974je}. This gives an additional contribution to the linear potential \eq{6.17}. An analogous homogeneous solution gives rise to a linear potential also in $D=3+1$ dimensions \cite{Hoyer:2021adf}.  

In the concluding Sec. \ref{sec7} I discuss the relation to previous solutions of the 't Hooft model, where the quarks have only positive kinetic energy. The negative energy components have important consequences. They induce an overlap between bound states, $\bra{B\,C}A\rangle$, suggestive of dual dynamics and color string breaking. Because the color electric field is of \order{g^2N_c} bound states thus acquire finite widths even in the $N_c \to \infty$ limit.

\section{Bound state equation in coordinate space} \label{sec2}  

\subsection{Invariance under time independent gauge transformations} \label{sec2A} 

Physical states should satisfy \eq{1.5} in temporal gauge \cite{Willemsen:1977fr,*Bjorken:1979hv,*Leibbrandt:1987qv,*Strocchi:2013awa}. This ensures that the states are invariant under time independent gauge transformations, which are generated by $\mG_a(t,x)$ \eq{1.4} and preserve the $A^0=0$ gauge condition. Denoting the infinitesimal gauge parameter by $\la_a(x)$, the unitary transformation generated by $\mG_a$
\begin{align} \label{2.1}
U_G(t)= 1+i\int dx\,\mG_a(t,x)\la_a(x) +\morder{\la_a^2}
\end{align} 
transforms the quark and gluon fields as
\begin{align} \label{2.2}
U_G(t)\psi(t,x)U_G^\dag(t) &= [1+ig\la_a(x)T^a]\psi(t,x) = \exp[ig\la_a(x)T^a]\psi(t,x) \equiv V_\la(x)\psi(t,x) \nn\crt
U_G(t)\bar\psi(t,x)U_G^\dag(t) &= \bar\psi(t,x)V_\la^\dag(x) \nn\crt
U_G(t)A_a(t,x)U_G^\dag(t) &= A_a(t,x) + \dx\la_a(x) + gf_{abc}A_b(t,x)\la_c(x)
\end{align}
At $t=0$ and with $U_G\ket{0}=\ket{0}$ \eq{1.5}, $U_G$ acts on the bound state \eq{1.6} as, 
\begin{align}
U_G\ket{M,P}&= \int dx_q dx_\qb\,\bar\psi(x_q)V_\la^\dag(x_q)U_G\,\mathbb{P}(x_\qb \to x_q)U_G^\dag\,\exp[\halft iP(x_q+x_\qb)]\Phip(x_q-x_\qb)V_\la(x_\qb)\psi(x_\qb)\ket{0}  \nn \crt
& =\ket{M,P}\hspace{1cm} \mbox{provided} \hspace{1cm}
U_G\,\mathbb{P}(x_\qb \to x_q)U_G^\dag =V_\la(x_q)\mathbb{P}(x_\qb \to x_q)V_\la^\dag(x_\qb) \label{2.3}
\end{align}
The last equality is a general feature of path-ordered exponentials as shown, \eg, in Sec. 15.3 of \cite{Peskin:1995ev}. I verify \eq{2.3} in Appendix \ref{appA}. Since $\mG(t,x)$ commutes with the Hamiltonian \cite{Willemsen:1977fr,*Bjorken:1979hv,*Leibbrandt:1987qv,*Strocchi:2013awa} the state will then satisfy $\mG_a(t,x)\ket{M,P;t}=0$ at all times $t$.

\subsection{Bound state momentum} \label{sec2B}  

The QCD$_2$ generators of space ($\hat P^1$) and time ($\hat P^0 = H$) translations as well as of boosts ($\hat M^{01}$) in temporal gauge are recalled in Appendix \ref{appB}. The momentum and energy operators \eq{b9} are
\begin{align}
\hat P^1(t) &= \int dx\big[E_a\rder_x A_a + \psi^\dag(-i\rder_x)\psi\big]  \label{2.4} \crt
\hat P^0(t) &= H(t) = \int dx\big[\halft(E_a)^2 +\psi^\dag\big(- i\alpha_1\rder_x +\gz m -g\alpha_1 A_a T^a \big)\psi\big] \hspace{2cm} (\alpha_1 \equiv \gz\go) \label{2.5}
\end{align}
The unitary operators $U_x(\ell)=\exp(-i\ell\hat P^1)$ and $U_t(\tau)=\exp(i\tau\hat P^0)$ translate the quark and gluon fields by $\ell$ in space and by $\tau$ in time, respectively \eq{b6}. Hence
\begin{align} \label{2.6}
U_x(\ell)\ket{M,P} 
&= \int dx_q dx_\qb\,\bar\psi(x_q+\ell)\mathbb{P}\exp\Big[ig\int_{x_\qb}^{x_q}dx\,A_a(x+\ell)T_a\Big]\,\exp[\halft iP(x_q+x_\qb)]\Phip(x_q-x_\qb)\psi(x_\qb+\ell)\ket{0} \nn\crt
&= \exp(-i\ell P)\ket{M,P}
\end{align}
where the state was brought to its standard form \eq{1.6} by a change of integration variables, $x \to x'=x+\ell$. This verifies that $\hat P^1\ket{M,P} = P\ket{M,P}$, \ie, that the state has momentum $P$.

\subsection{Bound state energy} \label{sec2C} 

The energy (Hamiltonian) eigenstate condition
\begin{align} \label{2.7}
\hat P^0\ket{M,P} = E_P\ket{M,P} \hspace{2cm} E_P = \sqrt{P^2+M^2}
\end{align}
imposes a BSE on the wave function $\Phip(x)$. Since the color electric field satisfies $E_a\ket{0}=0$ [Gauss' constraint \eq{1.5} does not involve particle creation],\footnote{This holds also in $D=3+1$ dimensions for the longitudinal electric field $\Ev_{L,a}$ in temporal gauge ($A_a^0 =0$).}
\begin{align} \label{2.8}
E_a^2(x)\bP(x_\qb \to x_q)\ket{0} = \big[E_a(x),\com{E_a(x)}{\bP(x_\qb \to x_q)}\big]\ket{0}
\end{align}
Using the discretization \eq{a1} and the canonical commutators \eq{1.3} for $x_\qb\leq x \leq x_q$,
\begin{align} \label{2.9}
&\bP(x_\qb \to x_q) = \bP(y_{l+1} \to x_q)\big[1+ig\Delta x A_b(y_l)T^b +\halft (ig\Delta x A_b(y_l)T^b)^2 +\morder{(\Delta x)^3}\big]\bP(x_\qb \to y_{l-1}) \hspace{1cm} (x_\qb<y_l<x_q) \nn\crt
&\com{E_a(y_{l})}{\bP(x_\qb \to x_q)}\ket{0} = \bP(y_{l+1}\to x_q)\big[i^2 g T^a+\halft i^3g^2\Delta x\big(T^a A_b(y_{l})T^b+A_b(y_{l})T^b T^a\big)\big]\bP(x_\qb \to y_{l-1})\ket{0} \nn\crt
&\halft\big[E_a(y_{l}),\com{E_a(y_{l})}{\bP(x_\qb \to x_q)}\big]\ket{0} = \halft\bP(y_{l+1} \to x_q)i^4 g^2 T^a T^a \bP(x_\qb \to y_{l-1})\ket{0} =\halft g^2C_F \bP(x_\qb \to x_q)\ket{0}
\end{align}
which is independent of $y_l$. The integral of $E_a^2(x)$ over $x$ in \eq{2.5} gives the linear potential \eq{1.7a} of QCD$_2$ \cite{tHooft:1974pnl},
\begin{align} \label{2.10}
\int_{x_\qb}^{x_q} dx\,\halft E_a^2(x)\bP(x_\qb \to x_q)\ket{0} = \halft g^2C_F|x_q-x_\qb|\bP(x_\qb \to x_q)\ket{0} \equiv V(x_q-x_\qb)\bP(x_\qb \to x_q)\ket{0}
\end{align}

The commutator of the quark part of the Hamiltonian \eq{2.5} contributes,
\begin{align} \label{2.11}
&\int dx\comb{\psi^\dag(x)(-i\alpha_1\rder_x +m\gz-g\alpha_1 A_a(x)T^a)\psi(x)}{\bar\psi_\alpha(x_q)\psi_\beta(x_\qb)}\ket{0} \nn\crt
&= \bar\psi(x_q)(-i\alpha_1\lder_{x_q} +m\gz+g\alpha_1 A_a(x_q)T^a)_\alpha \psi_\beta(x_\qb)\ket{0}
-\bar\psi_\alpha(x_q)\,{_\beta}(-i\alpha_1\rder_{x_\qb} +m\gz-g\alpha_1 A_a(x_\qb)T^a)\psi(x_\qb)\ket{0}
\end{align}
When partially integrating $-i\lder_{x_q}\to i\rder_{x_q}$ in $\ket{M,P}$ \eq{1.6} the derivative acts on $\bP(x_\qb \to x_q)$, on $\exp[iP(x_q+x_\qb)/2]$ and on $\Phip(x_q-x_\qb)$, similarly for $-i\rder_{x_\qb}\to i\lder_{x_\qb}$. The derivatives of $\bP(x_\qb \to x_q)$ cancel the $A_a$ terms as in \eq{a1},
\begin{align} \label{2.12}
\alpha_1\big[i\rder_{x_q}+gA_a(x_q)T^a\big]\bP(x_\qb\to x_q) = 0 \hspace{2cm}
\bP(x_\qb \to x_q)\big[i\lder_{x_\qb}-gA_a(x_\qb)T^a\big]\alpha_1 =0
\end{align}
The contribution from $-\bar\psi(x)g\alpha_1 A_a(x)T^a\psi(x)\ket{0}$ gives quark loops which are suppressed as $g \to 0$ in the $N_c \to\infty$ limit \cite{tHooft:1973alw,*Witten:1979kh}.
We are left with the bound state equation for the wave function,
\begin{align} \label{2.13}
i\dx\acom{\alpha_1}{\Phip(x)}-\halft P\com{\alpha_1}{\Phip(x)}+m\com{\gz}{\Phip(x)} +V(x)\Phip(x) = E_P\Phip(x)\,; \hspace{1.cm} V(x) = \halft g^2 C_F|x|
\end{align}
In the next section I summarize the main properties of $\Phip(x)$ as given by this BSE \cite{Dietrich:2012un,Hoyer:2021adf}.

\section{Properties of $\Phip(x)$} \label{sec3} 

\subsection{Parity and charge conjugation: $\mP\ket{M,P}=\eta_P\ket{M,-P},\ \mC\ket{M,P}=\eta_C\ket{M,P}$} \label{sec3A} 

From the expression \eq{1.6} of $\ket{M,P}$ and the action of the parity operator $\mP$ on the fields \eq{b15},
\begin{align} \label{3.1}
\mP\ket{M,P} &= \int dx_q dx_\qb\,\bar\psi(-x_q)\gz\mathbb{P}\exp\Big[-ig\int_{x_\qb}^{x_q}dx\,A_a(-x)T_a\Big]\,\exp[iP(x_q+x_\qb)/2]\Phip(x_q-x_\qb)\gz\psi(-x_\qb)\ket{0} \nn\crt
&= \int dx_q dx_\qb\,\bar\psi(x_q)\mathbb{P}\exp\Big[ig\int_{x_\qb}^{x_q}dx\,A_a(x)T_a\Big]\,\exp[-iP(x_q+x_\qb)/2]\gz\Phip(-x_q+x_\qb)\gz\psi(x_\qb)\ket{0} \nn\crt
&= \eta_P \ket{M,-P} \hspace{1cm} \mbox{if}\ \ \ \ \gz\Phip(-x)\gz = \eta_P\,\Phi^{\scriptscriptstyle(-P)}(x)
\end{align}
The parity transformation reverses the center of mass momentum $P$ of the state. From \eq{2.13} we see that $\gz\Phip(-x)\gz$ satisfies the BSE with $P \to -P$.

From the action of the charge conjugation operator $\mC$ on the fields \eq{b16},
\begin{align} \label{3.2}
\mC\ket{M,P} &= \int dx_q dx_\qb\,\psi^T(x_q)\go\mathbb{P}\exp\Big\{-ig\int_{x_\qb}^{x_q}dx\,[A_a(x)T_a]^T\Big\}\,\exp[iP(x_q+x_\qb)/2]\Phip(x_q-x_\qb)\go{\bar\psi(x_\qb)}^T\ket{0} \nn\crt
&=-\int dx_q dx_\qb\,\bar\psi(x_\qb)\mathbb{P}\exp\Big[-ig\int_{x_\qb}^{x_q}dx\,A_a(x)T_a\Big]\,\exp[iP(x_q+x_\qb)/2]\go\big[\Phip(x_q-x_\qb)\big]^T\go\psi(x_q)\ket{0} \nn\crt
&=\int dx_q dx_\qb\,\bar\psi(x_q)\mathbb{P}\exp\Big[-ig\int_{x_q}^{x_\qb}dx\,A_a(x)T_a\Big]\,\exp[iP(x_q+x_\qb)/2] i\go\big[\Phip(x_\qb-x_q)\big]^T i\go\psi(x_\qb)\ket{0} \nn\crt
&= \eta_C \ket{M,P} \hspace{1cm} \mbox{if}\ \ \ \ i\go\big[\Phip(-x)\big]^Ti\go = \eta_C\,\Phip(x)
\end{align}

Let us check that $\go\big[\Phip(-x)\big]^T\go$ satisfies the BSE \eq{2.13}. Taking its transpose and recalling that $\alpha_1^T=\alpha_1 = \gz\go$,
\begin{align} \label{3.3}
i\dx\acom{\alpha_1}{\big[\Phip(x)\big]^T}+\halft P\com{\alpha_1}{\big[\Phip(x)\big]^T}-m\com{\gz}{\big[\Phip(x)\big]^T} = (E_P-V)\big[\Phip(x)\big]^T
\end{align}
Changing $x\to -x$ and multiplying from both sides with $\go$ gives \eq{2.13} since $\acom{\go}{\alpha_1} = \acom{\go}{\gz} =0$,
\begin{align} \label{3.4}
i\dx\acom{\alpha_1}{\go\big[\Phip(-x)\big]^T\go}-\halft P\com{\alpha_1}{\go\big[\Phip(-x)\big]^T\go}+m\com{\gz}{\go\big[\Phip(-x)\big]^T\go} = (E_P-V)\go\big[\Phip(-x)\big]^T\go 
\end{align}

\vspace{-7mm}

This demonstrates that $\go\big[\Phip(-x)\big]^T\go$ satisfies the BSE if $\Phip(x)$ is a solution.

\subsection{Analytic solution of the bound state equation for $\Phip(x)$} \label{sec3B}

Expanding the $2\times 2$ Dirac wave function $\Phip_{\alpha\beta}(x)$ in scalar functions $\phip_j(x)\ (j=0,1,2,3)$,
\begin{align} \label{3.5}
\Phip(x) = \phip_0(x) +\alpha_1\,\phip_1(x) +\go\phip_2(x)+\gz\phip_3(x) \hspace{2cm}
(\alpha_1 = \sigma_1,\ \ \go = i\sigma_2,\ \ \gz = \sigma_3)
\end{align}
the parity \eq{3.1} and charge conjugation \eq{3.2} conditions imply,
\begin{align} \label{3.6}
\phip_{0,3}(-x) &= \eta_P\,\phi^{\scriptscriptstyle(-P)}_{0,3}(x) \hspace{2cm} 
\phip_{1,2}(-x) = -\eta_P\,\phi^{\scriptscriptstyle(-P)}_{1,2}(x)  \nn \crt
\phip_{0}(-x) &= \eta_C\,\phip_{0}(x) \hspace{2cm} 
\phip_{1,2,3}(-x) = -\eta_C\,\phip_{1,2,3}(x) 
\end{align}
At $P=0$ these require $\eta_P = \eta_C$, and are compatible since $\phi^{\scriptscriptstyle (P=0)}_3(x)=0$ \eq{3.7}. In the following I consider $\Phip(x \geq 0)$, with $\Phip(x < 0)$ given by \eq{3.6}.

Inserting the expansion \eq{3.5} of $\Phip(x\geq 0)$ into the BSE \eq{2.13} gives the following relations between the components,
\begin{align}
&\phip_2(x) = \frac{E_P-V}{(E_P-V)^2-P^2}\,2m\phip_1(x)
 \hspace{2cm} \phip_3(x) = \frac{P}{(E_P-V)^2-P^2}\,2m\phip_1(x)  \label{3.7}\crt
& \frac{2i}{E_P-V}\dx\phip_1(x) = \phip_0(x) \hspace{3cm} 
  \frac{2i}{E_P-V}\dx\phip_0(x) = \Big[1-\frac{4m^2}{(E_P-V)^2-P^2}\Big] \phip_1(x)  \label{3.8}
\end{align}
In terms of the variable $\tau_P(x)$ \eq{1.7b},
\begin{align} \label{3.9}
V'\tau_P(x) \equiv (E_P-V)^2-P^2  \hspace{2cm}
\dx = -2(E_P-V)\partial_{\tau_P} \hspace{2cm} (x \geq 0)
\end{align}
the relations \eq{3.8} become
\begin{align} \label{3.10}
\partial_{\tau_P}\phip_1(\tau_P) &= \frac{i}{4}\phip_0(\tau_P) \hspace{2cm}
\partial_{\tau_P}\phip_0(\tau_P) = \frac{i}{4}\Big(1-\frac{4m^2}{V'\tau_P}\Big)\phip_1(\tau_P) \nn\crt
&\partial_{\tau_P}^2 \phip_1(\tau_P) + \frac{1}{16}\Big(1-\frac{4m^2}{V'\tau_P}\Big)\phip_1(\tau_P) = 0
\end{align}
where $\phip(x) \equiv \phip[\tau_P(x)]$. Remarkably, relations \eq{3.10} do not explicitly depend on  $E_P$ or $P$. This holds only for a linear potential $V(x) = V'|x|$, which ensures the relation between $\dx$ and $\partial_{\tau_P}$ in \eq{3.9}. Hence the $P$-dependence of the wave functions $\phip_{0,1}[\tau_P(x)]$ as functions of $x$ is given by the $P$-dependence of $\tau_P(x)$ \eq{3.9}. The components $\phip_{2,3}(x)$ have an explicit dependence on $P$,
\begin{align} \label{3.11}
\phip_2(x) = \frac{E_P-V}{V'\tau_P(x)}\,2m\phip_1[\tau_P(x)]
 \hspace{2cm} \phip_3(x) = \frac{P}{V'\tau_P(x)}\,2m\phip_1[\tau_P(x)]
\end{align}

Defining the $x$-dependent ``boost parameter'' $\zeta_P(x)$ by
\begin{align} \label{3.12}
\cosh\zeta_P(x) = \frac{E_P-V(x)}{\sqrt{V'\tau_P}} \hspace{2cm} \sinh\zeta_P(x) =\frac{P}{\sqrt{V'\tau_P}}
\end{align}
the full wave function \eq{3.5} may be expressed as
\begin{align} \label{3.13}
\Phip(x) = \phi_0 + \phi_1\Big[\alpha_1+ \frac{2m(E_P-V)}{V'\tau_P}\,\go + \frac{2mP}{V'\tau_P}\,\gz\Big]
= e^{-\alpha_1\zeta_P/2}\Big(\phi_0 + \phi_1\alpha_1+\frac{2m}{\sqrt{V'\tau_P}}\phi_1\,\go\Big)e^{\alpha_1\zeta_P/2}
\end{align}
In the latter expression the term in (\ ) depends on $\tau_P$ only, whereas $\zeta_P$ depends also explicitly on $P$. 

The (arbitrarily normalized) solution of \eq{3.10} for $x \geq 0$ can be expressed in terms of the Kummer function \cite{Dietrich:2012un,Hoyer:2021adf},
\begin{align} \label{3.14}
\phip_1(\tau_P) &= V'\tau_P\,\exp(-i\tau_P/4)\kum(1-im^2/2V',2,i\tau_P/2) = \phi_1^*(\tau_P) \nn\crt
\phip_0(\tau_P) &= -\phi_1(\tau_P) -4iV'\,\exp(-i\tau_P/4)\kum(1-im^2/2V',1,i\tau_P/2) = -\phi_0^*(\tau_P)
\end{align}
This solution of the second-order differential equation \eq{3.10} was determined by the requirement $\phi_1(\tau_P=0)=0$, which ensures that $\phip_2(x)$ and $\phip_3(x)$ \eq{3.11} are regular at $\tau_P=0$. The symmetry $\phip_1(-x) = - \eta_C \phip_1(x)$ imposes a continuity condition at $x=0$, \ie, at $V'\tau_P(x=0)=E_P^2-P^2$. This value must be independent of $P$ for continuity to hold in all frames, requiring $E_P^2-P^2 = M^2$. Imposing continuity at any $P$ determines the allowed values of the bound state masses $M$.

The $P$-dependence of $\Phip(x)$ found here from the BSE \eq{2.13} should agree with that given by the boost generator $\hat M^{01}$. This was shown in Coulomb gauge in \cite{Dietrich:2012iy}, here I show it for temporal gauge in Appendix \ref{appC}. Even though concerning only infinitesimal boosts, both demonstrations are technically involved. A simpler method might provide better physical insight.

\subsection{Large separations $x$ between the quarks} \label{sec3C} 

The variable $\tau_P(x)$ \eq{3.9} is quadratically dependent on $x$. From $\tau_P(x=0)=M^2/V'$ it decreases to $\tau_P(x=E/V')=-P^2/V'$, then increases again, with $\tau_P(x\to\infty) \propto V'x^2$. At large $\tau_P$ the \order{1/\tau_P} term in the BSE \eq{3.10} may be neglected, so $\phip_1(\tau_P \to \infty) \propto \exp(\pm i\tau_P/4)$. The analytic solution \eq{3.14} gives
\begin{align} \label{3.15}
\lim_{|\tau_P|\to\infty}\big[\phi_1(\tau_P)+\phi_0(\tau_P)\big]&=\lim_{|\tau_P|\to\infty}\big[\phi_1(\tau_P)-\phi_0(\tau_P)\big]^* \nn\crt
&=N\exp\big[i\tau_P/4-i(m^2/2V')\log(|\tau_P|/2) + i\arg\Gamma(1+im^2/2V')-i\pi/2\big] \nn\crt
N&= \frac{4{(V')}^{3/2}}{\sqrt{\pi}\,m}\sqrt{\exp(\pi m^2/V')-1}\;e^{-\theta(-\tau)\pi m^2/2V'} \hspace{2cm} 
\end{align}
where $\theta(x) = 1\ (0)$ for $x >0\ (x<0)$. The oscillating behavior means that the wave function is not (globally) normalizable, and that the kinetic energy term $\partial_{\tau_P}^2\phip_1$ of the BSE \eq{3.10} is negative at large $\tau_P$. If the quark fields in the definition \eq{1.6} of the state are expanded in the free basis as in \eq{b13} the $b_k d_k\ket{0}$ component will dominate at large $x$. The other operator components of the state are normalizable.

It is helpful to recall an analogous feature of the Dirac equation for an electron in a linear potential. This was noted already in 1932 \cite{Plesset:1930zz}, see also Sec. IV of \cite{Hoyer:2021adf}. For a large potential ($V \gsim m$) the eigenstates of the Dirac equation include $e^+e^-$ pairs (in the free fermion basis). Perturbatively the pairs are due to time-ordered ``$Z$''-diagrams, and they give rise to the negative (kinetic) energy components of the wave function. If the Dirac state is expressed as (a superposition of) $\psi^\dag(x)\ket{0}$, the pairs reside in $\ket{0}$. A basis without pairs is obtained by Bogoliubov transforming the free creation and annihilation operators in $\psi^\dag(x)$. Then at small values of the potential (small $|x|$ for a linear potential) the single fermion is mostly an electron, while at large $|x|$ it is mostly a positron. This reveals the physics: A linear potential that confines electrons repels positrons. The Dirac wave function oscillates with phase $\propto x^2$ at large $|x|$, describing positrons with momenta $\propto x$. In analogy to plane waves, the wave function is not normalizable and the energy levels are continuous.

The $q\qb$ bound states \eq{1.6} considered here have a discrete mass spectrum, for which the wave function components $\phip_2(x)$ and $\phip_3(x)$ \eq{3.7} are regular at $\tau_P =0$. The negative energy components of the wave function (corresponding to contributions from the annihilation operators in $\bar\psi(x_q)$ and $\psi(x_\qb)$) allow a nonvanishing overlap $\bra{B\,C}A\rangle$, \ie, a ``fragmentation'' of bound state $A$ into $B+C$. Physically, it is akin to string breaking, \ie, the production/annihilation of a $q\qb$ pair by the bound state potential. Since the potential $V(x) = \halft g^2C_F |x|$ \eq{1.7a} is of leading order in the $N_c \to \infty$ limit, the fragmentation is not suppressed. The bound state overlap qualitatively resembles the dual features of hadron dynamics; initially produced (``parent'') states describe final (``daughter'') states in an average sense  \cite{Melnitchouk:2005zr}. These aspects merit further study.

I refer to \cite{Dietrich:2012un,Hoyer:2021adf} for further features of $\Phip(x)$. 

\section{Momentum space} \label{sec4} 

The wave function $\Phip(k)$ of momentum space is related to $\Phip(x)$ of coordinate space through a Fourier transform,
\begin{align} \label{4.1}
\Phip(x) &= \int_{-\infty}^{\infty}\frac{dk}{2\pi}\,e^{ikx}\, \Phip(k) \hspace{2cm} \Phip(k) = \int_{-\infty}^{\infty} dx\,e^{-ikx}\, \Phip(x)
\end{align}
The parity and charge conjugation symmetries \eq{3.1} and \eq{3.2} are in momentum space ($\eta_P = \eta_C$ in $D=1+1$),
\begin{align} \label{4.1b}
\gz\Phip(-k)\gz = \eta_P\,\Phi^{\scriptscriptstyle(-P)}(k) \hspace{2cm}
i\go\big[\Phip(-k)\big]^Ti\go = \eta_C\,\Phip(k)
\end{align} 
The quark $q$ and antiquark $\qb$ momenta are defined in terms of their relative momentum $k$ as,
\begin{align} \label{4.2}
q \equiv \halft P +k \hspace{2cm} \qb \equiv \halft P -k  \hspace{2cm} (-\infty < q,\qb < \infty)
\end{align}

\subsection{Energy projection} \label{sec4A} 

Kinetic energy projection is defined in terms of the free $u(k)$ and $v(k)$ spinors \eq{b14}. With $E_k \equiv \sqrt{k^2+m^2}$,
\begin{align} \label{4.3}
U_+(q) &= \frac{u(q)}{\sqrt{2E_q}} = \frac{E_q+m+\alpha_1\, q}{\sqrt{2E_q(E_q+m)}}\Big({\textstyle\begin{array}{c} 1 \\ 0 \end{array}} \Big) \hspace{2cm} 
U_-(q) = \frac{v(-q)}{\sqrt{2E_q}} = \frac{E_q+m-\alpha_1\, q}{\sqrt{2E_q(E_q+m)}}\Big({\textstyle\begin{array}{c} 0 \\ 1 \end{array}} \Big) \nn\crt
V_+(\qb) &= \frac{v(\qb)}{\sqrt{2E_\qb}} = \frac{E_\qb+m+\alpha_1\, \qb}{\sqrt{2E_\qb(E_\qb+m)}}\Big({\textstyle\begin{array}{c} 0 \\ 1 \end{array}} \Big) \hspace{2cm} 
V_-(\qb) = \frac{u(-\qb)}{\sqrt{2E_\qb}} = \frac{E_\qb+m-\alpha_1\, \qb}{\sqrt{2E_\qb(E_\qb+m)}}\Big({\textstyle\begin{array}{c} 1 \\ 0 \end{array}} \Big)
\end{align}
The projectors on positive and negative kinetic energy for the quark and antiquark are then, respectively: 
\begin{align} \label{4.4}
U_+(q)U_+^\dag(q) &= \inv{2E_{q}}(E_{q}+\gz m+\alpha_1 q) \hspace{2cm}
U_-(q)U_-^\dag(q)= \inv{2E_{q}}(E_{q}-\gz m-\alpha_1 q)\nn\crt
V_+(\qb)V_+^\dag(\qb) &= \inv{2E_{\qb}}(E_{\qb}-\gz m+\alpha_1 \qb) \hspace{2cm}
V_-(\qb)V_-^\dag(\qb) = \inv{2E_{\qb}}(E_{\qb}+\gz m-\alpha_1 \qb) \nn\crt
\sum_{s=\pm} U_s(q) U_s^\dag(q) &= \sum_{s=\pm} V_{s}(\qb) V_{s}^\dag(\qb) =1  \hspace{2.5cm} U_s^\dag(q)U_{s'}(q) = V_{s}^\dag(\qb)V_{s'}(\qb) = \delta_{s,s'}
\end{align}
The energy projected $(s,\sbb = \pm)$ wave functions $\phip_{s\bar s}(k)$ are defined by \eq{4.2} and
\begin{align} \label{4.5}
\phip_{s\bar s}(k) \equiv \bar U_s(q)\Phip(k)V_{\bar s}(\qb) \hspace{2cm}
\Phip(k)=\sum_{s,\bar s=\pm}\gz U_s(q)\phip_{s\bar s}(k)V_{\bar s}^\dag(\qb)
\end{align}

\subsection{BSE in momentum space} \label{secVB} 

The BSE \eq{2.13} is Fourier transformed and energy projected using the spinors \eq{4.3},
\begin{align} \label{4.6}
\int_{-\infty}^\infty dx\,e^{-ikx}\,\bar U_s(q)\big[{\rm Eq.\ }\eq{2.13}\big]V_\sbb(\qb) \hspace{2cm} q=\halft P+k,\ \ \ \qb=\halft P-k,\ \ \ \ s,\sbb = \pm
\end{align}

\subsubsection{Kinetic energies} \label{secVB1}

The quark kinetic energy term contributes,
\begin{align} \label{4.7}
\int dx\,e^{-ikx}&\,\bar U_s(q)\big[\alpha_1(i\rder_x-\halft P)+\gz m\big]\Phip(x)V_\sbb(\qb)
= U_s^\dag(q)(\alpha_1\, q + \gz m)\gz\Phip(k)V_\sbb(\qb) \nn\crt
&= E_q\,U_s^\dag(q)\big[2U_+(q)U_+^\dag(q)-1\big]\gz\,\Phip(k)V_\sbb(\qb)
= sE_q\,\bar U_s(q)\,\Phip(k)V_\sbb(\qb) = sE_q\,\phip_{s\sbb}(k)
\end{align}
The antiquark kinetic term similarly contributes
\begin{align} \label{4.8}
\bar U_s(q)\int dx&\,\Phip(x)\big[\alpha_1(i\lder_x+\halft P)-\gz m\big]e^{-ikx}\,V_\sbb(\qb)
= \bar U_s(q)\,\Phip(k)(\alpha_1\, \qb - \gz m)\,V_\sbb(\qb) \nn\crt
&= E_\qb\,\bar U_s(q)\,\Phip(k)\big[2V_+(\qb)V_+^\dag(\qb)-1\big]\,V_\sbb(\qb)
= \sbb E_\qb\,\bar U_s(q)\,\Phip(k)\,V_\sbb(\qb) = \sbb E_q\,\phip_{s\sbb}(k)
\end{align}
Including the remaining terms in \eq{2.13} the BSE becomes
\begin{align} \label{4.9}
\big(E_P-sE_q-\sbb E_\qb\big)\phip_{s\sbb}(k) = \bar U_s(q)\int dx\,e^{-ikx}\,V(x)\Phip(x)\,V_\sbb(\qb)
\end{align}
where $E_P = \sqrt{P^2+M^2}$ and the linear potential $V(x) = V'|x|$ with $V'= \halft g^2C_F = \quart g^2(N_c^2-1)/N_c$.

\subsubsection{Potential energy} \label{secVB2}  

Since the potential $V(x)=V'|x|$ is local in $x$ its contribution becomes a convolution in $k$. We have
\begin{align} \label{4.10}
V'|x|\Phip(x) = 2V'\int_{-\infty}^\infty \frac{dl'}{2\pi}\,\inv{l'^2}(1-e^{il'x})\int\frac{dk'}{2\pi}\,e^{ik'x}\Phip(k')
\end{align}
The $1/l'^2$ term regularizes the $l'$-integral, adding an (infinite) constant to $V(x)$ which in the BSE $\eq{4.9}$ may be absorbed in the energy $E_P$ of the state.
Changing variables $k'\to k,\ l'\to l-k$ in the $1/l'^2$ term, and $k'\to l,\ l'\to k-l$ in the $e^{il'x}/l'^2$ term,
\begin{align} \label{4.11}
V'|x|\Phip(x) = -2V'\int_{-\infty}^\infty \frac{dk}{2\pi}\,e^{ikx}\dashint_{-\infty}^\infty \frac{dl}{2\pi}\,\inv{(l-k)^2}\big[\Phip(l)-\Phip(k)\big]
\end{align}
where the $l$-integral is regularized by the principal value. The BSE \eq{4.9} becomes
\begin{align} \label{4.12}
\big(E_P-sE_q-\sbb E_\qb\big)\phip_{s\sbb}(k) = -2V'\dashint_{-\infty}^\infty \frac{dl}{2\pi}\,\inv{(l-k)^2}\big[\bar U_s(q)\Phip(l)\,V_\sbb(\qb)-\phip_{s\sbb}(k)\big]
\end{align}
Using \eq{4.5} and defining $p=\halft P+l,\  \pb=\halft P-l$ we have 
\begin{align} \label{4.13}
\bar U_s(q)\Phip(l)V_{\sbb}(\qb) &= \sum_{s',\sbb'=\pm}U_s^\dag(q)U_{s'}(p)\,\phip_{s'\sbb'}(l)\,V_{\sbb'}^\dag(\pb) V_{\sbb}(\qb) \equiv \sum_{s',\sbb'=\pm}\Theta_{ss'}(q,p)\,\phip_{s'\sbb'}(l)\,\Theta_{\sbb'\sbb}(\pb,\qb) \nn\crt
\Theta_{ss'}(q,p)  &\equiv  U_s^\dag(q)U_{s'}(p) \hspace{2cm} 
V_{\sbb'}^\dag(\pb)V_{\sbb}(\qb) = \Theta_{\sbb'\sbb}(\pb,\qb)
\end{align}
From \eq{4.3} the expressions for $\Theta_{ss'}(q,p)$ are
\begin{align} \label{4.14}
\Theta_{++}(q,p) &= \frac{(E_q+m)(E_p+m)+qp}{\sqrt{4E_qE_p(E_q+m)(E_p+m)}} \hspace{2cm} 
\Theta_{+-}(q,p) = \frac{(E_p+m)q-(E_q+m)p}{\sqrt{4E_qE_p(E_q+m)(E_p+m)}} \nn\crt
\Theta_{-+}(q,p) &= -\Theta_{+-}(q,p)  \hspace{4.5cm} \Theta_{--}(q,p) = \Theta_{++}(q,p)
\end{align}
The BSE in momentum space is then
\begin{align} \label{4.15}
(E_P-sE_{q}-\sbb E_{\qb})\phip_{s\sbb}(k) =-2V'\dashint_{-\infty}^\infty \frac{dl}{2\pi}\,\inv{(l-k)^2}\Big[ \sum_{s',\sbb'=\pm} \Theta_{ss'}(q,p)\phip_{s'\sbb'}(l)\Theta_{\sbb'\sbb}(\pb,\qb)-\phip_{s\sbb}(k)\Big]
\end{align}
The $\Theta$-functions determine the mixing of the positive and negative energy components of the wave function caused by the potential. Because 
\begin{align} \label{4.16}
\Theta_{ss'}(q,p)\Big|_{l=k}=\delta_{ss'} 
\end{align}
the numerator of the integrand is of \order{l-k} near $l=k$, and the integral is defined by its principal value.

\section{Infinite momentum frame}  \label{sec5}

\subsection{The wave function for $P \to \infty$}  \label{sec5A}  

In the limit of $P \to \infty$ at fixed $\tau_P(x)$ \eq{1.7b} the quark separation $x = x_q-x_\qb$ Lorentz contracts. The IMF variable $\taui(u)$ becomes a linear function of the scaled separation $u \equiv E_P x \simeq Px$.
\begin{align} \label{5.1}
\tau_P(x) &=  M^2/V'-2E_P |x| + V'x^2\ \  \to\ \  \taui(x)= M^2/V'-2E_P |x| = M^2/V'-2|u| 
\end{align}
When the wave function $\Phip(x) \to \Phii(u)$ is expanded into Dirac components as in \eq{3.5}, the $\phii_2(u) \simeq \phii_3(u)$ components \eq{3.7} dominate $\phii_0(u)$ and $\phii_1(u)$ \eq{3.8} by one power of $P$. Recalling also the charge conjugation symmetry \eq{4.1b} and the analytic expression \eq{3.14},
\begin{align} \label{5.2}
\Phii(u) & \equiv 2mP\,(\gz+\go)\phii(u) \nn\crt
\phii(u) &= \exp\big[-i\quart \taui(u)\big]\kum\big[1-im^2/2V',2,i\halft\taui(u)\big]
= -\eta_C \phii(-u) \hspace{1cm} (u\geq 0)
\end{align}

In the IMF the quark and antiquark momenta \eq{4.2} are defined in terms of the momentum fraction $\xb$,
\begin{align} \label{5.3}
q = \xb P \hspace{2cm} \qb = (1-\xb) P \hspace{2cm} k= (\xb-\halft)P  \hspace{2cm} (-\infty < \xb < \infty)
\end{align}
Note that since $k$ ranges over all positive and negative values at any $P$ \eq{4.1}, the same is true for $\xb$. In physical processes such as deep inelastic scattering the range $0<\xb<1$ is mandated by positive final state energy, but this does not constrain wave functions. The Fourier transform \eq{4.1} becomes in the IMF,
\begin{align} \label{5.4}
\Phii(\xb) &= \int_{-\infty}^\infty \frac{du}{P}\, e^{-iuk/P}\,\Phii(u) = 2m(\gz+\go)\int_{-\infty}^\infty du\,e^{-iu(\xb-1/2)}\phii(u) \equiv 2m(\gz+\go)\phii(\xb) \nn\crt
\phii(\xb) &= \int_0^\infty du\,\phii(u)\big\{\exp[-i(\xb-\halft)u] -\eta_C\,\exp[i(\xb-\halft)u]\big\}
= -\eta_C\,\phii(1-\xb)
\end{align}
The charge conjugation symmetry \eq{3.6} allowed to restrict the integral to $u \ge 0$, for which the analytic expression \eq{5.2} of $\phii(u)$ applies. Using the integral representation,
\begin{align} \label{5.5}
_1F_1(a,b,z) = \frac{1}{\Gamma(a)\,\Gamma(b-a)}\int_0^1 dt\,e^{zt}t^{a-1}(1-t)^{b-a-1}
\end{align}
where
\begin{align} \label{5.6}
z &= i\taui(u)/2 = iM^2/2V'-iu  \nn\crt
\frac{1}{\Gamma(a)\,\Gamma(b-a)} &= \frac{1}{\Gamma(1-im^2/2V')\,\Gamma(1+im^2/2V')}   
= \frac{2V'}{\pi m^2} \sinh\Big(\frac{\pi m^2}{2V'}\Big) \equiv N_m
\end{align}
\begin{align} \label{5.7}
\phii(\xb) &= N_m \int_0^\infty du\,e^{-iM^2/4V' + iu/2}\int_0^1 dt\,e^{itM^2/2V'-itu}\Big(\frac{1-t}{t}\Big)^{im^2/2V'}\Big[e^{-i(\xb-1/2)u} -\eta_C\,e^{i(\xb-1/2)u}\Big] \nn\crt
&= N_m\int_0^1 dt\,e^{i(t-1/2)M^2/2V'}\Big(\frac{1-t}{t}\Big)^{im^2/2V'}\int_0^\infty du\,
\Big[e^{i(1-\xb-t)u} -\eta_C\,e^{i(\xb-t)u}\Big]
\end{align}
The integral over $u$ is defined by shifting $\xb \to \xb-i\epsilon$ ($\xb \to \xb+i\epsilon$) in the first  (second) term,
\begin{align} \label{5.8}
\phii(\xb) &= N_m \int_0^1 dt\,e^{i(t-1/2)M^2/2V'}\Big(\frac{1-t}{t}\Big)^{im^2/2V'}
\Big(\frac{i}{1-\xb-t+i\epsilon} -\eta_C\,\frac{i}{\xb-t+i\epsilon}\Big)
=-\eta_C\, {\phii(\xb)}^*
\end{align}
where $-\infty<\xb<\infty$. The $t$-integral is regular for all $\xb \neq 0$ or 1. In the limit of $\xb \to 1$ the behavior of the integral is governed by the phase $-\log(t)m^2/2V'$. This gives rise to terms $\propto (1-\xb)^{\pm im^2/2V'}$, with a finite norm but a logarithmically divergent phase. The same holds for $\xb \to 0$.

\begin{figure} \centering
\includegraphics[width=1.\columnwidth]{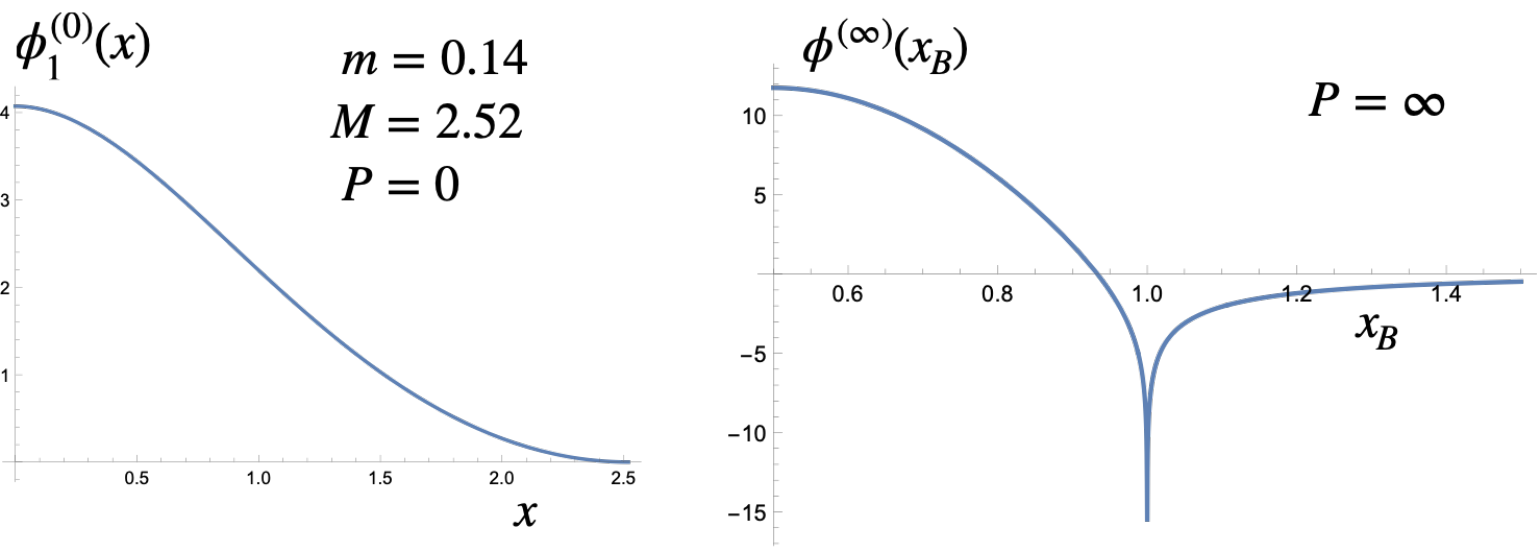}
\caption{Features of the $\eta_C=-1$ ground state, with quark mass $m=0.14$ and bound state mass $M= 2.51954$ $(V'=1)$.\\ \textit{Left:} Plot of rest frame wave function $\phi_1^{\scriptscriptstyle(P=0)}(x)$ \eq{3.14} for $0\leq x \leq M$. $\phir_1(x\to\infty)$ oscillates according to \eq{3.15}.\\ \textit{Right:} The IMF wave function $\phii(\xb)=\phii(1-\xb)$ \eq{5.8} for $0.5 \leq \xb \leq 1.5$. $\phii(\xb \to \infty) \simeq -0.40/x_B^2$. \label{f1}}
\end{figure}

As a numerical illustration I consider the $\eta_C=-1$ ground state with quark mass $m=0.14 \sqrt{V'}$ and bound state mass $M= 2.51954 \sqrt{V'}$. The component $\phi_1^{\scriptscriptstyle(P=0)}(x)=\phir_1(-x)$ of the the rest frame wave function $\Phir(x)$ \eq{3.5} is shown in \fig{f1} (left) as a function of the quark separation $x=x_q - x_\qb$. The condition $\dx\phir_1(0)=0$ ensures continuity at $x=0$, while $\phir_1(M/V')=0$ is required for $\phir_2(M/V')$ and $\phir_3(M/V')$ to be finite \eq{3.11}. These conditions and $m$ determine the bound state mass $M$. At large quark separations $\phir_1(|x| \to\infty)$ oscillates with constant norm according to \eq{1.7b} and \eq{3.14} \cite{Dietrich:2012un,Hoyer:2021adf}. 

\fig{f1} (right) shows the IMF $(P\to\infty)$ wave function $\phii(\xb)=\phii(1-\xb)$ \eq{5.8} as a function of the quark momentum fraction $\xb = q/P$. The quark momentum ranges over $-\infty< q < \infty$ at any $P$ \eq{4.2}, and thus also $-\infty< \xb < \infty$. The deep inelastic scattering (DIS) cross section is described by $|\phii(0<\xb <1)|^2$, with $\xb$ identified as the Bjorken variable \cite{Dietrich:2012un,Hoyer:2021adf}. In analogy to Dirac wave functions the strong potential generates also negative kinetic energy components of $\phii(\xb)$ at $\xb<0$ and $\xb>1$ (see below).

\subsection{Energy projection in the IMF}  \label{sec5B}  

The spinors \eq{4.3} simplify in the $P\to\infty$ limit. With $\veps_q = 1\ (-1)$ for $q> 0\ (q<0)$,
\begin{align} \label{5.9}
\Ui_+(q) = \inv{\sqrt{2}} \Big({\textstyle\begin{array}{c} 1 \\ \veps_q \end{array}} \Big)& \hspace{1cm} 
\Ui_-(q) = \inv{\sqrt{2}} \Big({\textstyle\begin{array}{c} -\veps_q \\ 1 \end{array}} \Big) \hspace{1cm} 
\Vi_+(\qb) = \inv{\sqrt{2}} \Big({\textstyle\begin{array}{c} \veps_\qb \\ 1 \end{array}} \Big) \hspace{1cm} 
\Vi_-(\qb) = \inv{\sqrt{2}} \Big({\textstyle\begin{array}{c} 1 \\ -\veps_\qb \end{array}} \Big) \nn\crt
{\Ui_s}^\dag(q)(1+\alpha_1) &= (1+s\veps_q)(1\ \ 1)\inv{\sqrt{2}} \hspace{3cm} 
(1+\alpha_1)\Vi_\sbb(\qb) = \inv{\sqrt{2}} \Big({\textstyle\begin{array}{c} 1 \\ 1 \end{array}} \Big)(1+\sbb\veps_\qb)
\end{align}

The energy projected ($s,\sbb=\pm$) wave function defined in \eq{4.5} and \eq{5.2} becomes, with $\phi^{\scriptscriptstyle (P\to\infty)}_{s\sbb}(k) \equiv \phii_{s\sbb}(\xb)$,
\begin{align} \label{5.10}
\phii_{s\sbb}(\xb) &\equiv \inv{2m}\,\bar U_s^{\scriptscriptstyle{(\infty)}}(q) \Phii(\xb) \Vi_{\sbb}(\bar q)
= {\Ui_s}^\dag(p)(1+\alpha_1)\Vi_\sbb(\bar p)\,\phii(\xb) 
= \halft(1+s\,\veps_q)(1+\sbb\,\veps_{\bar q})\phii(\xb)
\end{align}
where $\veps_q = \veps_\xb$ since $q=\xb P$, and similarly $\veps_\qb = \veps_{1-\xb}$. Hence for $0<\xb<1$ only $\phii_{++}(\xb)$ is nonvanishing, while for $\xb<0$ only $\phii_{-+}(\xb)$ and for $\xb>1$ only $\phii_{+-}(\xb)$ are nonzero. The $\phii_{--}(\xb)$ component vanishes for all $\xb$ in the $P\to\infty$ limit at fixed $\tau_P$. 

On the lhs. of the BSE \eq{4.15} we may approximate $E_P = \sqrt{P^2+M^2} \simeq P + M^2/2P$, and similarly for $E_q$ and $E_\qb$,
\begin{align} \label{5.11}
(E-s E_{q}-\sbb E_{\qb})\phii_{s\sbb}(\xb) \simeq \Big[(1-s|\xb|-\sbb|(1-\xb)|)P+ \inv{2P}\Big(M^2-s\,\frac{m^2}{|\xb|} -\sbb\,\frac{m^2}{|1-\xb|}\Big)\Big]\phii_{s\sbb}(\xb)
\end{align}  
The form of $\phii_{s\sbb}(\xb)$ \eq{5.10} ensures that the \order{P} term vanishes in the $P \to \infty$ limit for all $s,\sbb=\pm$.
The rhs. of the BSE \eq{4.15} is, with $p=\yb P$, $l = (\yb-\halft)P$ and $\phi^{\scriptscriptstyle (P\to\infty)}_{s\sbb}(l) \equiv \phii_{s\sbb}(\yb) $,
\begin{align} \label{5.12}
-\frac{V'}{\pi P}\dashint_{-\infty}^\infty \frac{d\yb}{(\yb-\xb)^2}\Big[ \sum_{s',\sbb'=\pm} \Ti_{ss'}(q,p)\phii_{s'\sbb'}(\yb)\Ti_{\sbb'\sbb}(\pb,\qb)-\phii_{s\sbb}(\xb)\Big]
\end{align}

The factors $\Ti_{ss'}(q,p)={\Ui_s}^\dag(q)\Ui_{s'}(p)$ of \eq{4.14} and \eq{5.9} in the IMF are,
\begin{align} \label{5.13}
\Ti_{++}(q,p) &= \halft(1+\veps_q\veps_p) \hspace{2.1cm} \Ti_{+-}(q,p) = \halft(\veps_q-\veps_p) \nn\crt
\Ti_{-+}(q,p) &=-\Ti_{+-}(q,p)  \hspace{2cm} \Ti_{--}(q,p) =\Ti_{++}(q,p) 
\end{align}

For $0<\xb<1$ and $s=\sbb=+$ we have in \eq{5.11} $\phii_{++}(\xb)= 2\phii(\xb)$ \eq{5.10}. Since $q>0$ and $\qb>0$ the first term in \eq{5.12} reduces to $2\phii(\yb)$ for all $\yb$,
\begin{align} \label{5.14}
\hspace{-.3cm}\sum_{s',\sbb'=\pm} \Ti_{+s'}(q,p)\phii_{s'\sbb'}(\yb)\Ti_{\sbb'+}(\pb,\qb) =
\left\{{\textstyle\begin{array}{l} 
\Ti_{+-}(q>0,p<0)\phii_{-+}(\yb<0)\Ti_{++}(\pb>0,\qb>0) =2\phii(\yb<0) \crt 
\Ti_{++}(q>0,p>0)\phii_{++}(0<\yb<1)\Ti_{++}(\pb>0,\qb>0) =2\phii(0<\yb<1) \crt
\Ti_{++}(q>0,p>0)\phii_{+-}(\yb>1)\Ti_{-+}(\pb<0,\qb>0) =2\phii(\yb>1)
\end{array}}\right.
\end{align}
Consequently the BSE for the IMF wave function $\phii(\xb)$ of \eq{5.8} is, for $0<\xb<1$ and $s=\sbb=+$,
\begin{align} \label{5.15}
&\Big(M^2-\frac{m^2}{\xb} -\frac{m^2}{1-\xb}\Big)\phii(\xb) = -\frac{2V'}{\pi}\dashint_{-\infty}^\infty \frac{d\yb}{(\yb-\xb)^2}\,[\phii(\yb)-\phii(\xb)]
\end{align}
where $V' = \halft g^2C_F$. This may be compared with Eq. (25) of 't Hooft \cite{tHooft:1974pnl}, which multiplied with $g^2/\pi$ reads
\begin{align} \label{5.16}
\Big(M^2-\frac{m^2}{x} -\frac{m^2}{1-x}\Big)\phi(x) = -\frac{g^2}{\pi}\Big(\inv{x}+\inv{1-x}\Big)\phi(x) - \frac{g^2}{\pi}\dashint_0^1 \frac{\phi(y)}{(y-x)^2}\,dy
= - \frac{g^2}{\pi}\dashint_0^1 \frac{\phi(y)-\phi(x)}{(y-x)^2}\,dy
\end{align}
The quark propagator corrections, which give rise to the terms $-(g^2/\pi)[1/x+1/(1-x)]\phi(x)$ in \eq{5.16}, decrease the singularity of the integral at $y=x$. There are no quark propagators in present approach, where the less singular form of the integral is a consequence of the Fourier transform \eq{4.11}.
Identifying 't Hooft's coupling $g^2 \to g^2C_F = 2V'$, $x \to \xb$ and $y \to \yb$ \eq{5.16} agrees with \eq{5.15}, \textit{except} that the negative energy components $\phii_{-+}(y_B<0)$ and $\phii_{+-}(y_B>1)$ extend the integration over $\yb$ from $[0,1]$ to $[-\infty,\infty]$.

The BSE \eq{5.15} actually holds for all $\xb\neq 0,1$, \ie, $-\infty < \xb < \infty$. For $\xb < 0,\ s=-,\ \sbb=+$,
\begin{align} \label{5.17}
\hspace{-.3cm}\sum_{s',\sbb'=\pm} \Ti_{-s'}(q,p)\phii_{s'\sbb'}(\yb)\Ti_{\sbb'+}(\pb,\qb) =
\left\{{\textstyle\begin{array}{l} 
\Ti_{--}(q<0,p<0)\phii_{-+}(\yb<0)\Ti_{++}(\pb>0,\qb>0) =2\phii(\yb<0) \crt 
\Ti_{-+}(q<0,p>0)\phii_{++}(0<\yb<1)\Ti_{++}(\pb>0,\qb>0) =2\phii(0<\yb<1) \crt
\Ti_{-+}(q<0,p>0)\phii_{+-}(\yb>1)\Ti_{-+}(\pb<0,\qb>0) =2\phii(\yb>1)
\end{array}}\right.
\end{align}
Due to the symmetry $\phii(\xb)=-\eta_C\, \phii(1-\xb)$ \eq{5.4} the BSE \eq{5.15} is unchanged when $\xb \to 1-\xb$. The singular behavior of the lhs. as $\xb\to 1$ is matched by a singularity on the rhs., since $\lim_{\epsilon\to 0}\phii(\yb=1-\epsilon) \neq \lim_{\epsilon\to 0}\phii(\yb=1+\epsilon)$. Table \ref{t1} demonstrates the validity of the BSE for the $\eta_C=-1$ ground state of \fig{f1}.

\begin{table} \caption{Numerical results for the BSE \eq{5.15} at various values of $\xb$, for the bound state of \fig{f1}. The second to fourth columns show the contributions to the right hand side from the indicated integration ranges over $\yb$. The full result for the rhs (fifth column) agrees with the lhs of \eq{5.15} (last column).} \vspace{.3cm} \label{t1}
\begin{tabular}{|r||c|c|c|c|c|}\hline $\ \xb\ $&$\ -\infty<\yb<0\ $&$\ 0 < \yb < 1\ $&$\ 1 <\yb < \infty\ $&$\ -\infty<\yb<\infty\ $&lhs of \eq{5.15}\\ \hline\hline 
$-0.2\ $&4.671&$-12.096$&$-0.328$&$-7.754$&$-7.754$\\ \hline
$0.5\ $&16.338&41.025&16.335&\ 73.698\ \ &\ 73.698\ \ \\ \hline
$0.8\ $&5.475&7.168&25.291&\ 37.934\ &\ 37.934\ \\ \hline
$0.9\ $&1.810&$-19.004$&28.440&\ 11.246\ &\ 11.246\ \\ \hline
$0.99\ $&$-3.051$&$-70.332$&49.712&\ $-23.672$\ &\ $-23.672$\ \\ \hline
$\ 0.999\ $&$-6.043$&$-113.155$&253.917&\ 134.719\ &\ 134.719\ \\ \hline
$1.2\ $&$-0.328$&$-12.096$&4.671&\ $-7.754$\ \ &\ $-7.754$\ \ \\ \hline
$1.5\ $&0.010&$-4.759$&1.727&\ $-3.022$\ \ &\ $-3.022$\ \ \\ \hline
\end{tabular}
\end{table}

\section{Using the temporal gauge constraint to determine $E_a$}  \label{sec6}

\subsection{Reproducing the linear potential}  \label{sec6A}

The eigenvalue condition $H\ket{M,P} = E\ket{M,P}$ for the states $\ket{M,P}$ \eq{1.6} of momentum $P$,
\begin{align} \label{6.1}
\ket{M,P}= \int dx_q dx_\qb\,\bar\psi(x_q)\mathbb{P}\exp\Big[ig\int_{x_\qb}^{x_q}dx\,A_a(x)T_a\Big]\,\exp[iP(x_q+x_\qb)/2]\Phip(x_q-x_\qb)\psi(x_\qb)\ket{0}
\end{align}
determines the BSE \eq{2.13} for the wave function $\Phip(x_q-x_\qb)$. In Section \ref{sec2C} the Hamiltonian $\hat P^0=H$ \eq{2.5} was determined canonically from the QCD$_2$ action in temporal ($A^0=0$) gauge, 
\begin{align} \label{6.2}
H = \int dx\,\big[\halft (E_a)^2 + \psi^\dag(-i\alpha_1\rder_x+\gz m-gA_a T^a\alpha_1)\psi\big]
\end{align}
Physical states are required to be invariant under time-independent gauge transformations, which preserve $A^0=0$. Those transformations are generated by Gauss' operator \eq{1.4},
\begin{align} \label{6.3}
\mathcal{G}_a(t,x) \equiv \frac{\delta\mS_{QCD}}{\delta{A_a^0(t,x)}} = \dx E_a(t,x)+g f_{abc}A_b E_c(t,x)-g\psi^\dag T^a\psi(t,x)
\end{align}
In Sec. \ref{sec2A} and Appendix \ref{appA} I verified that $\mathcal{G}_a(x)\ket{M,P}=0$. Hence
\begin{align} \label{6.4}
\dx E_a(x)\ket{M,P} = g\big[- f_{abc}A_b E_c(x)+\psi^\dag T^a\psi(x)\big]\ket{M,P}
\equiv g\,\mE_a(x)\ket{M,P}
\end{align}
Viewed as a differential equation for $E_a(x)$ this provides an alternative expression for the $E_a^2$ contribution to $H$. Recalling that $\dx^2 |x-y| = 2\delta(x-y)$ we may solve for $E_a(x)$,
\begin{align} \label{6.5}
E_a(x) = \halft g\, \dx \int dy\, |x-y|\,\mE_a(y)
\end{align}
The contribution of the color electric field to the Hamiltonian $H$ \eq{6.2} is thus
\begin{align} \label{6.6}
H_V \equiv \halft\int dx\, E_a^2(x) &= \frac{g^2}{8}\int dx \Big[\dx\int dy\, |x-y|\,\mE_a(y)\Big]
\Big[\dx\int dz\, |x-x|\,\mE_a(z)\Big] \nn\crt
 &= -\frac{g^2}{4} \int dy\,dz\,|y-z|\,\mE_a(y)\,\mE_a(z)
\end{align}

The fermion part $\psi^\dag T^a\psi(x)$ of $\mE_a(x)$ acting on the quark fields in $\ket{M,P}$ \eq{6.1} gives,
\begin{align} \label{6.7}
\com{\mE_a(x)}{\bar\psi_A(x_q)} = \bar\psi_{A'}(x_q)T^a_{A'A} \delta(x-x_q) \hspace{2cm}
\com{\mE_a(x)}{\psi_A(x_\qb)} = -T^a_{AA'}\psi_{A'}(x_\qb) \delta(x-x_\qb)
\end{align}
The $- f_{abc}A_b E_c(x)$ term in $\mE_a(x)$ \eq{6.4} operating on the gauge link in $\ket{M,P}$ is suppressed by \order{g} compared to the fermion contribution. Ignoring the gauge link contribution we have for a (globally) color singlet $q\qb$ state,
\begin{align} \label{6.8}
\ket{q(x_q)\qb(x_\qb)} &\equiv \sum_A\bar\psi_A^\alpha(x_q)\psi_A^\beta(x_\qb)\ket{0} \crt
H_V\ket{q(x_q)\qb(x_\qb)} &= \halft g^2 |x_q-x_\qb| \bar\psi(x_q)T^a T^a \psi(x_\qb)\ket{0} = \halft g^2 C_F |x_q-x_\qb| \ket{q(x_q)\qb(x_\qb)} = V(x_q-x_\qb)\ket{q(x_q)\qb(x_\qb)} \nn
\end{align}
The potential $V(x)=\halft g^2C_F|x|$ agrees with that previously obtained in \eq{2.10}, where it arose from the commutator of $E_a^2$ with the gauge link in $\ket{M,P}$.

\subsection{Adding a homogeneous solution to $E_a$}  \label{sec6B}

The electric fields generated by QED and QCD bound states differ in an important respect. The electric field $E(x)$ of QED$_2$ is given by \eq{6.5} without a gauge boson or color matrix $T^a$ contribution,
\begin{align} \label{6.9}
E(x) = \halft e\, \dx \int dy\, |x-y|\,\psi_e^\dag(y)\psi_e(y) =  \halft e\int dy\, \veps(x-y)\,\psi_e^\dag(y)\psi_e(y)
\end{align}
where $\veps(x)= 1\ (-1)$ for $x>1\ (x<1)$.
The electric field of an $e^-e^+$ state $\ket{e^-(x_e)e^+(x_\eb)} \equiv \bar\psi_e^\alpha(x_e)\psi_e^\beta(x_\eb)\ket{0}$ is
\begin{align} \label{6.10}
E(x)\ket{e^-(x_e)e^+(x_\eb)} &= \halft e\big[\veps(x-x_e)-\veps(x-x_\eb)\big]\ket{e^-(x_e)e^+(x_\eb)}
\end{align}
The field is nonvanishing only between the charges. For $x_e>x_\eb$,
\begin{align} \label{6.11}
\frac{\bra{e^-(x_e)e^+(x_\eb)}E(x)\ket{e^-(x_e)e^+(x_\eb)}}{\bra{e^-(x_e)e^+(x_\eb)}e^-(x_e)e^+(x_\eb)\rangle}
= \left\{\begin{array}{l} -e\ \ \ (x_\eb<x<x_e) \\ \ 0\ \ \ \ \ \mbox{otherwise} \end{array} \right.
\end{align}
In QCD we have instead, from \eq{6.5}  and \eq{6.7},
\begin{align} \label{6.12}
\bra{q(x_q)\qb(x_\qb)}E_a(x)\ket{q(x_q)\qb(x_\qb)} \propto \tr T^a =0 \hspace{1cm} \mbox{for\ all}\ x
\end{align}
External observers see a vanishing color field due to the sum over quark colors. Equivalently, a color singlet state does not give rise to a color octet field. However, there is no sum over colors in the binding of a meson's $q\qb$ constituents. A red quark feels the color octet field generated by its anti-red anti-quark companion in the Fock state, and \textit{vice versa}. 

In $D=3+1$ dimensions Gauss' law determines the (dipole) electric field of an $e^+e^-$ state. The solution is unique for fields that vanish at large distances, which is required to avoid long range interactions in QED. For color singlet $q\bar q$ states there is no such requirement, since the state does not generate a color octet field at any $\xv$.

This motivates adding a homogeneous solution to \eq{6.5}, for which $\dx E_a(x)=0$,
\begin{align} \label{6.13}
E_a(x) =  \dx \int dy\big[-\kappa\, x\,y+ \halft g\, |x-y|\big]\mE_a(y)
\end{align}
The normalization $\kappa$ may depend on the state $\ket{M,P}$ that $E_a(x)$ acts on. Linearity in $x$ ensures that the homogeneous (sourceless) part of $E_a(x)$ is independent of $x$. Linearity in $y$ ensures translation invariance for color singlet states. After partial integrations $H_V = \halft \int dx\, E_a^2(x)$ becomes
\begin{align} \label{6.14}
H_V = \int dy\,dz\big[yz\big(\halft\kappa^2\intt dx + g\kappa\big) -\quart g^2 |y-z|\big]\mE_a(y)\mE_a(z)
\end{align}
Acting on the state in \eq{6.8} gives
\begin{align} \label{e7.15}
H_V\ket{q(x_q)\qb(x_\qb)} = \big[(\halft\kappa^2\intt dx + g\kappa)C_F(x_q-x_\qb)^2+ \halft g^2C_F |x_q-x_\qb|\big]\ket{q(x_q)\qb(x_\qb)}
\end{align}
The coefficient of $\kappa^2$ is proportional to the volume of space: the homogeneous contribution has introduced an isotropic vacuum field energy density $E_\la$. This energy is irrelevant provided it is universal, \ie, the same for all states. Consequently we must take $\kappa^2$ to be inversely proportional to $(x_q-x_\qb)^2$. Defining the universal constant $\la$ by
\begin{align} \label{6.16}
\kappa \equiv \frac{\la^2}{gC_F}\,\inv{|x_q-x_\qb|} \hspace{2cm} E_\la = \frac{\la^4}{2g^2C_F}
\end{align}
gives, after subtracting the (infinite but universal) vacuum energy $E_\la\int dx$,
\begin{align} \label{6.17}
H_V\ket{q(x_q)\qb(x_\qb)} = \big(\la^2+\halft g^2C_F\big)|x_q-x_\qb|\ket{q(x_q)\qb(x_\qb)}
\end{align}
Inclusion of the homogeneous solution to $E_a(x)$ \eq{6.13} added $\la^2$ to the slope $V'$ of the linear potential. 

The possibility to include a homogeneous solution is potentially interesting in $D=3+1$ dimensions. It adds a linear term $V'|\xv|$ to the $-\as C_F/|\xv|$ Coulomb $q\qb$ potential, with $V'$ related (as above) to the energy density of the vacuum. The corresponding potentials for color singlet $qqq$, $gg$ and $q\qb g$ Fock states are also confining, and are correctly related to each other when the coordinates of two constituents coincide \cite{Hoyer:2021adf}.

\section{Discussion}  \label{sec7}

Color confinement is a central feature of hadron dynamics. Numerical (lattice) calculations have determined that the hadron spectrum of QCD agrees with data \cite{Kronfeld:2012uk}, whereas analytic methods remain elusive. Confinement has not been demonstrated using the formally exact Bethe-Salpeter \cite{Salpeter:1951sz} approach to bound states. QCD$_2$ may provide insights since the Coulomb potential is confining in $D=1+1$ dimensions. 't Hooft found \cite{tHooft:1974pnl} that the Bethe-Salpeter equation of QCD$_2$ can be solved exactly in the $N_c \to \infty$ limit, which inspired many further studies.

The present study of the 't Hooft model does not use the Bethe-Salpeter approach. This avoids quark propagators, which were found to be singular and to lack Lorentz covariance \cite{Bars:1977ud}. Instead, I directly determine the color singlet $\ket{q\qb}$ eigenstates \eq{1.6} of the QCD$_2$ Hamiltonian \eq{2.5}. Poincar\'e covariance emerges in an explicit, yet nontrivial way for these  equal-time states. The 't Hooft equation \eq{5.16} is reproduced \eq{5.15} with a surprising twist: The wave function $\phii(\xb)$ \eq{5.8} has negative energy components even in the IMF. Hence $\phii(\xb)$ does not vanish for $\xb\to 0,1$ as postulated by 't Hooft (\fig{f1}, right).

Negative energy solutions are familiar from the Dirac equation, but it may seem surprising that they survive in the IMF. Due to Lorentz contraction the potential is of \order{1/P} while the kinetic energies are of \order{P}. However, the kinetic energy contributions to the BSE cancel at \order{P} \eq{5.11}, leaving only \order{1/P} terms in the 't Hooft equation. A bound state constituent with negative kinetic energy in the rest frame is boosted to a large negative energy in the IMF. This allows the energy and momentum $\xb P$ of a quark to be balanced by that of an antiquark with $(1-\xb)P$ even if $\xb >1$. 

The $q\qb$ wave function \eq{3.15} oscillates $\propto \exp(iV'x^2/4)$ at large values of the variable $\tau_P(x)=M^2/V'-2E_P|x|+V'x^2$ \eq{1.7b}, and is thus not globally normalizable. There is a nonvanishing overlap $\bra{B\,C}A\rangle$ between bound states $A,B$ and $C$, due to $q\qb$ production in the binding potential, with features expected of ``string breaking''. When $E_A=E_B+E_C$ this allows the decay $A\to B+C$, giving the bound states finite widths of \order{g^2N_c}. It has previously been assumed that decays are due to the interaction term of \order{g}, implying zero widths at leading order in $N_c \to \infty$ \cite{tHooft:1973alw,*Witten:1979kh}. For $E_A\ll E_B+E_C$ on the other hand the rapid oscillations of the wave function should suppress contributions to physical processes. Further study is required to establish whether these features can give an explanation of the duality observed in data on hadron dynamics \cite{Melnitchouk:2005zr}.

The present bound state method suggests a mechanism for confinement in $D=3+1$ dimensions. In Sec. \ref{sec6B} I noted that adding a homogeneous term \eq{6.13} to the solution of Gauss' constraint \eq{1.5} implied an isotropic gluon field energy density $E_\la$ given by a new parameter $\la$ \eq{6.16}. This does not change the equations of motion and maintains full Poincar\'e covariance. Instead of taking the $N_c \to \infty$ limit we may now set $N_c=3$ and consider $\halft g^2N_c \ll \la^2$ in \eq{6.17}. This allows a perturbative expansion in $g^2$, which at lowest order gives all the previous results vith $V'=\la^2$. 

Adding a homogeneous solution similarly introduces a scale $\la_{QCD}$ and a confining instantaneous potential in $D=3+1$ \cite{Hoyer:2021adf}. Such a change of boundary condition for the gauge-dependent, longitudinal color electric field seems not to violate the basic principles of QCD. At \order{\as^0} the hadronic states and processes given by $\la_{QCD}$ qualitatively agree with data. The essential and nontrivial requirement of Poincar\'e covariance was verified for the electromagnetic form factor in \cite{Hoyer:2023noz}. The spontaneous breaking of chiral symmetry required for a quantitative phenomenology of light hadrons remains to be studied. 

\begin{acknowledgments}
I thank Matti J\"arvinen for helpful discussions and comments on the manuscript. I am privileged to be associated as Professor Emeritus to the Physics Department of the University of Helsinki.
\end{acknowledgments}

 \appendix
\renewcommand{\theequation}{\thesection.\arabic{equation}}

\section{Gauge invariance of the bound states \label{appA}}

Here I verify \eq{2.3}, $U_G\,\mathbb{P}(x_\qb \to x_q)U_G^\dag =V(x_q)\mathbb{P}(x_\qb \to x_q)V^\dag(x_\qb)$,
by discretizing in $x$ ($t=0$ is understood). For $x_\qb\leq x\leq x_q$ and $0 \leq n \leq N_\mathbb{P}$,
\begin{align} \label{a1}
x \to y_n &= x_\qb+n\Delta{x} \hspace{2cm} \Delta{x} = \frac{x_q-x_\qb}{N_\mathbb{P}} \hspace{2cm}
\int_{x_\qb}^{x_q} dx\,f(x) \to \Delta{x}\sum_{n=0}^{N_\mathbb{P}} f(y_n) \nn\crt
\bP(x_\qb \to x_q) &\equiv \mathbb{P}\exp\Big[ig\int_{x_\qb}^{x_q}dx\,A_a(x)T_a\Big] 
= \prod_{n=0}^{N_\mathbb{P}} \exp\big[ig\Delta{x\,A_a(y_n)T^a}\big]  \hspace{1cm} (N_\mathbb{P} \to \infty) \nn\crt
&=\prod_{n=l+1}^{N_\mathbb{P}} \exp\big[ig\Delta{x\,A_a(y_n)T^a}\big]\Big[1+\Delta x igA_a(y_l)T^a+\morder{(\Delta x)^2}\Big]\prod_{n=0}^{l-1} \exp\big[ig\Delta{x\,A_a(x_n)T^a}\big]  \nn\crt
&=\bP(y_l \to x_q)\big[1+\Delta x igA_a(y_l)T^a+\morder{(\Delta x)^2}\big]\bP(x_\qb \to y_l)
\hspace{1cm} (x_\qb< y_l < x_q) \nn\crt
\frac{\partial}{\partial x_q}\bP(x_\qb \to x_q) &= igA_a(x_q)T^a\bP(x_\qb \to x_q) \hspace{1cm}
\frac{\partial}{\partial x_\qb}\bP(x_\qb \to x_q) = -\bP(x_\qb \to x_q)\,igA_a(x_\qb)T^a
\end{align}
The discretized derivative and the commutation relations \eq{1.3} are,
\begin{align} 
\frac{\partial}{\partial{y_k}} f(y_k) &= \frac{f(y_k)-f(y_{k-1})}{\Delta x} \label{a2} \crt
\com{E_a(y_k)}{A_b(y_l)} = \frac{i}{\Delta x}\delta_{k,l}\delta_{a,b}& \hspace{2cm}
\acomb{\psi^\dag_{\alpha,A}(y_k)}{\psi_{\beta,B}(y_l)} = \delta_{\alpha,\beta}\delta_{A,B}\frac{\delta_{k.l}}{\Delta x}
\label{a3}
\end{align}
The generator of time-independent gauge transformations $\mG_a(t,x)$ is given in \eq{1.4}, the unitary transformation $U_G(t)$ in \eq{2.1} and the unitary matrix $V_\la(x)$ in \eq{2.2}.
For $x_\qb< y_l,y_{l+1} < x_q$, using \eq{a1} and up to \order{\Delta x} corrections,
\begin{align} \label{a4}
&\hspace{-1cm}\com{\frac{\partial}{\partial{y_l}}E_a(y_l)}{\bP(x_\qb \to x_q)} = \nn\crt
&=\inv{\Delta x}\bP(y_{l+2} \to x_q)\com{E_a(y_{l+1})-E_a(y_{l})}{\big(1+ig\Delta{x}A_b(y_{l+1})T^b\big)\big(1+ig\Delta{x}A_c(y_{l})T^c\big)}\bP(x_\qb \to y_{l-1}) \nn\crt
&= \inv{\Delta x}\bP(y_{l+2} \to x_q)\big\{i^2gT^a\big(1+ig\Delta{x}A_b(y_{l})T^b\big)-
\big(1+ig\Delta{x}A_b(y_{l+1})T^b\big)i^2gT^a\big\}\bP(x_\qb \to y_{l-1}) \nn\crt
&= \bP(y_{l} \to x_q)i^3g^2 A_b(y_l)(T^aT^b-T^bT^a)\bP(x_\qb \to y_{l}) 
=\bP(y_l \to x_q)\, g^2 f_{abc}A_b(y_l)T^c \,\bP(x_\qb \to y_l)\nn\crt
&\hspace{-1cm}\com{g f_{abc}A_b E_c(y_l)}{{\bP(x_\qb \to x_q)}}= \bP(y_l \to x_q)i^2g^2 f_{abc}A_b(y_l)T^c\,\bP(x_\qb \to y_l)
\end{align}
The last two lines cancel in $\mG_a(x)$,
\begin{align} \label{a5}
\com{\mG_a(x)}{\bP(x_\qb \to x_q)}=0 \hspace{1cm} \mathrm{for}\ \ x_\qb< x < x_q)
\end{align}
At the endpoints $y_l=x_q$ and $y_l=x_\qb$ the derivative of the commutator is discontinuous,
\begin{align} \label{a6}
\partial_{x_q}E_a(x_q) \equiv \frac{E_a(x_q+\Delta x)-E_a(x_q)}{\Delta x} \hspace{2cm}
\partial_{x_\qb}E_a(x_\qb) \equiv \frac{E_a(x_\qb)-E_a(x_\qb-\Delta x)}{\Delta x}
\end{align}
and $\com{E_a(x_q+\Delta x)}{\bP(x_\qb \to x_q)}=\com{E_a(x_\qb-\Delta x)}{\bP(x_\qb \to x_q)} =0$. Hence
\begin{align} \label{a7}
\com{\partial_{x_q}E_a(x_q)}{\bP(x_\qb \to x_q)} &= -\inv{\Delta x}\com{E_a(x_q)}{\bP(x_\qb \to x_q)}
= -\frac{i^2 g}{\Delta x}T^a\bP(x_\qb \to x_q) \nn\crt
\com{\partial_{x_\qb}E_a(x_\qb)}{\bP(x_\qb \to x_q)} &= \inv{\Delta x}\com{E_a(x_\qb)}{\bP(x_\qb \to x_q)}
= \frac{i^2 g}{\Delta x}\bP(x_\qb \to x_q) T^a
\end{align}
Analogously $\com{gf_{abc}A_bE_c(x_q+\Delta x/2)}{\bP(x_\qb \to x_q)}=\com{gf_{abc}A_bE_c(x_\qb-\Delta x/2)}{\bP(x_\qb \to x_q)} =0$. Hence \eq{2.3} is verified in the continuum limit,
\begin{align} \label{a8}
\com{\mG_a(x)}{\bP(x_\qb \to x_q)} &= \delta(x-x_q)gT^a\,\bP(x_\qb \to x_q)-\delta(x-x_\qb)\,\bP(x_\qb \to x_q)gT^a \nn\crt
U_G\,\bP(x_\qb \to x_q)U_G^\dag &= \bP(x_\qb \to x_q) + i\int dx\,\com{\mG_a(x)}{\bP(x_\qb \to x_q)}\la_a(x) \nn\crt
&= [1+igT^a\la_a(x_q)]\bP(x_\qb \to x_q)[1-igT^a\la_a(x_\qb)] = V_\la(x_q)\bP(x_\qb \to x_q)V_\la^\dag(x_\qb)
\end{align}

\section{Poincar\'e covariance \label{appB}}

\subsection{General definitions} \label{appB1} 

In $D=1+1$ dimensions we may use a two-dimensional Dirac algebra with Pauli matrices
\begin{align} \label{b1}
\gz = \sigma_3 \hspace{2cm} \go = i\sigma_2 \hspace{2cm} \gz\go \equiv \alpha_1 = \sigma_1
\end{align}
The unitary Poincar\'e operators are related to their generators as
\begin{align}
U_x(\ell) &= \exp(-i\ell\hat P^1) \ \ \ \ \ \mbox{Space translation by\ } \ell  \label{b2} \crt
U_t(\tau) &= \exp(i\tau\hat P^0) \ \ \ \ \ \ \ \mbox{Time translation by\ } \tau   \label{b3} \crt
U_{tx}(\xi) &= \exp(-i\xi\hat M^{01}) \ \ \ \mbox{Boost by\ } \xi   \label{b4}
\end{align}
The generators satisfy the equal-time Lie algebra
\beqa\label{b5}
 \com{\hat P^0}{\hat P^1}= 0 , \label{plie01}\  \hspace{2cm}
 \com{\hat P^0}{\hat M^{01}} = i \hat P^1 , \  \hspace{2cm}
 \com{\hat P^1}{\hat M^{01}} = i \hat P^0 
\eeqa
Infinitesimal translations of a field $\hat\mO$ are generated by
\begin{align}  \label{b6}
\comb{\hat P^1}{\hat{\cal{O}}}&=i\dx\hat{\cal{O}}: \hspace{1.3cm} U_x(\delta{x})\hat{\cal{O}}U_x^\dag(\delta{x})
= \hat{\cal{O}}-i\delta{x}\comb{\hat P^1}{\hat{\cal{O}}} = \hat{\cal{O}}+\delta{x}\,\dx\hat{\cal{O}} = \hat{\cal{O}}(x+\delta{x}) \nn\crt
\comb{\hat P^0}{\hat{\cal{O}}}&=-i\dt\hat{\cal{O}}:\hspace{1cm} U_t(\delta{t})\hat{\cal{O}}U_t^\dag(\delta{t}) 
= \hat{\cal{O}}+i\delta{t}\comb{\hat P^0}{\hat{\cal{O}}} = \hat{\cal{O}}+\delta{t}\,\dt\hat{\cal{O}} = \hat{\cal{O}}(t+\delta{t})
\end{align}

The Lie algebra ensures that the momentum and energy of a state $\ket{M,P}$, which satisfies $\hat P^1\ket{M,P} = P\ket{M,P}$ and $\hat P^0\ket{M,P} = E_P\ket{M,P}$, transform correctly under boosts,
\begin{align} \label{b7}
\hat P^1 (1-i\delta{\xi}\hat{M}^{01})\ket{M,P}&= \big\{(1-i\delta{\xi}\hat{M}^{01})\hat{P^1}-i\delta{\xi}\comb{\hat{P^1}}{\hat{M}^{01}}\big\}\ket{M,P} =
(P+\delta{\xi}E_P) (1-i\delta{\xi}\hat{M}^{01})\ket{M,P} \nn\crt
\hat P^0 (1-i\delta{\xi}\hat{M}^{01})\ket{M,P}&= \big\{(1-i\delta{\xi}\hat{M}^{01})\hat{P^0}-i\delta{\xi}\comb{\hat{P^0}}{\hat{M}^{01}}\big\}\ket{M,P} =
(E_P+\delta{\xi}P) (1-i\delta{\xi}\hat{M}^{01})\ket{M,P}
\end{align}
Hence we may identify $U_{tx}(\delta\xi)\ket{M,P} = \ket{M,P+\delta{\xi}E_P}$.

\subsection{Generators of QCD$_2$ in temporal gauge} \label{appB2}

The translation invariance of the action defines the conserved energy-momentum tensor (see, \eg, Secs. 7.3 and 7.4 of \cite{Weinberg:1995mt}). For a Lagrangian $\mL$ [here given by \eq{1.1}] depending on fields $\hat{\mO}^\ell$ (here $A_a \equiv A_a^1$ and $\psi$),
\begin{align} \label{b8}
T^\nu_{\ \mu} &\equiv \frac{\partial\mL}{\partial(\partial_\nu\hat{\mO}^\ell)}\,\partial_\mu\hat{\mO}^\ell -g^\nu_\mu\,\mL \hspace{2cm} \partial_\nu T^\nu_{\ \mu}=0 
\end{align}
The generators of space and time translations are time independent. With $\alpha_1 \equiv \gz\go$,
\begin{align} \label{b9}
\hat P^1 &\equiv \int dx\, T^{01} = -\int dx\,\frac{\partial\mL_{QCD}}{\partial(\partial_0\hat{\mO}^\ell)}\partial_1 \hat{\mO}^\ell 
=\int dx\big[E_a\rder_x A_a + \psi^\dag(-i\rder_x)\psi\big] \nn\crt
\hat P^0 = H &\equiv \int dx\, T^{00} = \int dx\Big[\frac{\partial\mL_{QCD}}{\partial(\partial_0\hat{\mO}^\ell)}\partial_0 \hat{\mO}^\ell -\mL\Big]
= \int dx\Big\{\halft(E_a)^2 +\psi^\dag\big(- i\alpha_1\rder_x +\gz m -g\alpha_1 A_a T^a \big)\psi\Big\}
\end{align}
The boost generator $\hat{M}^{01}$ \eq{b4} may be expressed in terms of $\hat P^1$ and the (Hermitian) Hamiltonian density $\mH$ in \eq{b9},
\begin{align} \label{b10}
\hat{M}^{01} &= t\hat{P}^1 - \int dx\,x\,\mH
= \int dx\Big\{t\big[E_a\rder_x A_a + \psi^\dag(-i\rder_x) \psi\big] 
-x \big[\halft (E_a)^2 + \psi^\dag(- i\alpha_1\lrder_x+\gz m -g\alpha_1 A_a T^a)\psi\big]\Big\} \nn\crt
&= \int dx\Big\{t\big[E_a\rder_x A_a + \psi^\dag(-i\rder_x) \psi\big]
-x \big[\halft (E_a)^2 + \psi^\dag(- i\alpha_1\rder_x+\gz m -g\alpha_1 A_a T^a +\halft i\alpha_1)\psi\big]\Big\}
\end{align}
where on the first line $\lrder = \halft (\rder-\lder)$ acts only on the fermion fields.

The Lie algebra \eq{b5} may be verified using the canonical commutation relations \eq{1.3}. More generally, all generators are independent of time, and the translation operators satisfy \eq{b6}. Then
\begin{align} \label{b11}
\comb{\hat P^0}{\hat P^1} &= -i\dt\hat P^1 =0 \nn\crt
\comb{\hat P^1}{\hat M^{01}} &= -\int dx\, x\, i\dx \mH = \int dx\, i\mH = i\hat P^0
\end{align}
The explicit time dependence of the boost generator $\hat M^{01}$ in \eq{b10} implies,
\begin{align} \label{b12}
\frac{d}{dt}\hat M^{01} &= \dt^{expl} \hat M^{01} +i\comb{\hat P^0}{\hat M^{01}}=0  \nn \crt
\comb{\hat P^0}{\hat M^{01}} &= i\dt^{expl} \hat M^{01} =i\hat P^1
\end{align}

\subsection{Parity and charge conjugation} \label{appB3} 

The fermion field may be expanded in the basis given by the free creation/annihilation operators,
\begin{align} 
\psi(t,x) &= \int \frac{dk}{2\pi\, 2E_k}\big[u(k)e^{-itE_k+ixk}\,b_k +v(k)e^{itE_k-ixk}\,d^\dag_k \big] \label{b13} \crt
u(k) &= \inv{\sqrt{E_k+m}}(E_k+m+\alpha_1 k)\Big({\textstyle\begin{array}{c} 1 \\ 0 \end{array}} \Big) \hspace{3.2cm} E_k = \sqrt{k^2+m^2} \nn\crt
v(k) &= \inv{\sqrt{E_k+m}}(E_k+m+\alpha_1 k)\Big({\textstyle\begin{array}{c} 0 \\ 1 \end{array}} \Big) = \alpha_1\, u(k) \hspace{1.7cm} \alpha_1 = \gz\go = \sigma_1 \label{b14}
\end{align}
The parity transformation $(t,x) \to (t,-x)$ is implemented by the operator $\mP$, which leaves the action \eq{1.1} invariant,
\begin{align} \label{b15}
&\mP\,b_k\,\mP^\dag = b_k \hspace{4cm} \mP\,d_k\,\mP^\dag = -d_k \nn\crt
\mP\,\psi(t,x)\,\mP^\dag = \gz \psi(t,-x)& \hspace{2cm} \mP\,\bar\psi(t,x)\,\mP^\dag = \bar\psi(t,-x) \gz   \hspace{2cm} \mP\,A^1_a(t,x)\,\mP^\dag = - A^1_a(t,-x)
\end{align}
For charge conjugation we have correspondingly, with $\go = \sigma_3\sigma_1 = i\sigma_2$,
\begin{align} \label{b16}
&\mC\,b_k\,\mC^\dag = d_k \hspace{4cm} \mC\,d_k\,\mC^\dag = b_k \nn\crt
\mC\,\psi(t,x)\,\mC^\dag = -\go {\bar\psi}^T(t,x)& \hspace{1.5cm} \mC\,\bar\psi(t,x)\,\mC^\dag = - {\psi}^T(t,x)\go \hspace{1.5cm} \mC\,A^1_a(t,x)T^a\,\mC^\dag = - \big[A^1_a(t,x)T^a\big]^T
\end{align}

\section{$P$-dependence of $\Phip$ from $\hat M^{01}\ket{M,P}$} \label{appC} 

Here I verify that the $P$-dependence of the wave function determined by the BSE \eq{2.13} agrees with that given by the boost generator $\hat M^{01}$ \eq{b10}.
Taking \eq{1.6} to define $\ket{M,P}$ at $t=0$ I set $t=0$ in the expression \eq{b10} for $\hat M^{01}$,
\begin{align} \label{c1}
\hat M^{01}(t=0) =\int dx\,\Big\{ -x\halft(E_a)^2 +\psi^\dag\big[-x(-i\alpha_1\rder_x +m\gz-gA_aT^a\alpha_1) +i\halft\alpha_1\big]\psi\Big\}
\end{align} 
The generator transforms the operators in $\ket{M,P}$ as follows,
\begin{align} \label{c2}
\com{\hat M^{01}}{\bar\psi_\alpha(x_q)} &= \bar\psi(x_q)\big\{-x_q(-i\alpha_1\lder_{x_q}+m\gz+\alpha_1 gA_a(x_q)T^a)-\halft i\alpha_1\big\}_\alpha \nn\crt
\com{\hat M^{01}}{\psi_\beta(x_\qb)} &= {_\beta\big\{}x_\qb(-i\alpha_1\rder_{x_\qb}+m\gz-\alpha_1 gA_a(x_\qb)T^a)-\halft i\alpha_1\big\}\psi(x_\qb) \crt
\com{\hat M^{01}}{\bP(x_\qb\to x_q)} &=\int dx\com{-\halft xE_a^2}{\bP(x_\qb\to x_q)}
= -\halft\int_{x_\qb}^{x_q}dx\,xg^2C_F\,\bP(x_\qb\to x_q) =-\halft(x_q^2-x_\qb^2)V'\,\bP(x_\qb\to x_q)\nn
\end{align}
where I used \eq{2.9} on the last line, and denoted $V'\equiv \halft g^2C_F$ \eq{1.7a}. Altogether,
\begin{align} \label{c3}
\hat M^{01}\ket{M,P} = \int &dx_qdx_\qb\,\bar\psi(x_q)\Big\{\nn\crt
&\big[-x_q(i\alpha_1\rder_{x_q}+m\gz+\alpha_1 gA_a(x_q)T^a)-\halft i\alpha_1\big]
\mathbb{P}(x_\qb\to x_q)\,\exp[iP(x_q+x_\qb)/2]\Phip(x_q-x_\qb)\nn\crt
&-\halft(x_q^2-x_\qb^2)V'\,\bP(x_\qb\to x_q)\,\exp[iP(x_q+x_\qb)/2]\Phip(x_q-x_\qb)\nn\crt
&+\mathbb{P}(x_\qb\to x_q)\,\exp[iP(x_q+x_\qb)/2]\Phip(x_q-x_\qb)\big[x_\qb(i\alpha_1\lder_{x_\qb}+m\gz-\alpha_1 gA_a(x_\qb)T^a)-\halft i\alpha_1\big]\nn\crt
&\hspace{10cm}\Big\}\psi(x_\qb)\ket{0}
\end{align}
As in \eq{2.12} the derivatives acting on $\mathbb{P}(x_\qb\to x_q)$ are cancelled by the $\alpha_1 gA_aT^a$ terms. For $\lder_{x_\qb}$ acting on $x_\qb$,
\begin{align} \label{c4}
-\halft i\alpha_1\Phip +\Phip\big[i\alpha_1(x_\qb\lder_{x_\qb})-\halft i\alpha_1\big]= -\halft i\com{\alpha_1}{\Phip} 
\end{align}
With these simplifications,
\begin{align} 
\hat M^{01}\ket{M,P} &= \int dx_qdx_\qb\,\bar\psi(x_q)\,\bP(x_\qb\to x_q)\Big\{\nn\crt
&-x_q(i\alpha_1\rder_{x_q}+m\gz)\,\exp[iP(x_q+x_\qb)/2]\Phip(x_q-x_\qb)\nn\crt
&-\halft(x_q^2-x_\qb^2)V'\,\exp[iP(x_q+x_\qb)/2]\Phip(x_q-x_\qb) -\halft i\com{\alpha_1}{\Phip}\nn\crt
&+\exp[iP(x_q+x_\qb)/2]\Phip(x_q-x_\qb)(i\alpha_1\lder_{x_\qb}+m\gz)x_\qb
\Big\}\psi(x_\qb)\ket{0} \label{c5} \crt
& \hspace{-1cm} = \int dx_qdx_\qb\,\bar\psi(x_q)\,\bP(x_\qb\to x_q)\,\exp[iP(x_q+x_\qb)/2]\Big\{&\nn\crt
&-ix_q\alpha_1\rder_{x_q}\Phip(x_q-x_\qb)-ix_\qb\rder_{x_q}\Phip(x_q-x_\qb)\alpha_1
+\halft x_qP\alpha_1\Phip -\halft x_\qb P\Phip\alpha_1 \nn\crt 
&-x_q m\gz\Phip+x_\qb m\Phip\gz
-\halft V'(x_q-x_\qb)(x_q+x_\qb)\, \Phip -\halft i\com{\alpha_1}{\Phip}
\Big\}\psi(x_\qb)\ket{0} \label{c6} 
\end{align}
With $x \equiv x_q-x_\qb$, replace
\begin{align} \label{c7}
x_q = \halft(x_q+x_\qb)+\halft x \hspace{2cm} x_\qb = \halft(x_q+x_\qb)-\halft x \hspace{2cm}
\partial_{x_q}\Phip(x_q-x_\qb) = \dx\Phip(x)
\end{align}
Use the bound state equation \eq{2.13} on the terms $\propto x_q+x_\qb$,
\begin{align} \label{c8}
-\halft(x_q+x_\qb)\Big(i\dx\acom{\alpha_1}{\Phip(x)}-\halft P\com{\alpha_1}{\Phip(x)}+m\com{\gz}{\Phip(x)}+V'x\Phip(x)\Big) = -\halft(x_q+x_\qb)E_P\Phip(x)
\end{align}
giving
\begin{align} \label{c9}
\hat M^{01}\ket{M,P} &= \int dx_qdx_\qb\,\bar\psi(x_q)\,\bP(x_q,x_\qb)\,\exp[iP(x_q+x_\qb)/2]\Big\{ \nn\crt
&-\halft(x_q+x_\qb)E_P\Phip-\halft ix\dx\com{\alpha_1}{\Phip} +\quart xP\acom{\alpha_1}{\Phip}
-\halft xm \acom{\gz}{\Phip} -\halft i\com{\alpha_1}{\Phip}\Big\}\psi(x_\qb)\ket{0} 
\end{align}
According to Eq. (A.93) of \cite{Hoyer:2021adf} the BSE \eq{2.13} implies
\begin{align} \label{c10}
-i\halft x\dx\com{\alpha_1}{\Phip}+\quart xP\acom{\alpha_1}{\Phip}-\halft xm\acom{\gz}{\Phip}
= \frac{ixP}{E_P-V}\dx\Phip -\frac{iV'x}{2(E_P-V)}\com{\alpha_1}{\Phip}
\end{align}
Use this in \eq{c9} to obtain
\begin{align} \label{c11}
\hat M^{01}\ket{M,P} = \int dx_qdx_\qb\,&\bar\psi(x_q)\,\bP(x_q,x_\qb)\,\exp[iP(x_q+x_\qb)/2]\Big\{ \nn\crt
&-\halft(x_q+x_\qb)E_P\Phip+\frac{ixP}{E_P-V}\dx\Phip -\frac{iE_P}{2(E_P-V)}\com{\alpha_1}{\Phip}\Big\}\psi(x_\qb)\ket{0}
\end{align}

From the definition of $\hat M^{01}$ in \eq{b4}, using $\partial_\xi P = E_P$,
\begin{align} 
U_{tx}(\delta{\xi})\ket{M,P} =&(1-i\delta{\xi}\hat M^{01})\ket{M,P}  \label{c12} \crt
-i \hat M^{01}\ket{M,P} =\frac{\delta}{\delta{\xi}}\ket{M,P} 
=& \frac{\delta}{\delta{\xi}}\int dx_q dx_\qb\,\bar\psi(x_q)\mathbb{P}\exp\Big[ig\int_{x_\qb}^{x_q}dx\,A_a(x)T_a\Big]\,\exp[\halft iP(x_q+x_\qb)]\Phip(x_q-x_\qb)\psi(x_\qb)\ket{0} \nn\crt
= \int dx_q dx_\qb\,\bar\psi(x_q)\mathbb{P}\exp&\Big[ig\int_{x_\qb}^{x_q}dx\,A_a(x)T_a\Big]\,\exp[iP(x_q+x_\qb)/2]\Big\{\halft iE_P(x_q+x_\qb)\Phip+\partial_{\xi}\Phip\Big\}\psi(x_\qb)\ket{0} \label{c13}
\end{align}
Comparing with \eq{c11} implies that $\hat M^{01}$ generates the following change of $\Phip(x)$,
\begin{align} \label{c14}
\partial_\xi\Phip(x) = \frac{xP}{E_P-V}\dx\Phip -\frac{E_P}{2(E_P-V)}\com{\alpha_1}{\Phip}
\end{align}
This agrees with Eq. (7.32) of \cite{Hoyer:2021adf}, which was derived from the $P$-dependence of the solution of the BSE. Hence the $P$-dependence induced by the boost agrees with that implied by the BSE \eq{2.13}, confirming the analysis in \cite{Dietrich:2012iy}.

\bibliography{refs.bib}

\begin{thebibliography}{29}%
\makeatletter
\providecommand \@ifxundefined [1]{%
 \@ifx{#1\undefined}
}%
\providecommand \@ifnum [1]{%
 \ifnum #1\expandafter \@firstoftwo
 \else \expandafter \@secondoftwo
 \fi
}%
\providecommand \@ifx [1]{%
 \ifx #1\expandafter \@firstoftwo
 \else \expandafter \@secondoftwo
 \fi
}%
\providecommand \natexlab [1]{#1}%
\providecommand \enquote  [1]{``#1''}%
\providecommand \bibnamefont  [1]{#1}%
\providecommand \bibfnamefont [1]{#1}%
\providecommand \citenamefont [1]{#1}%
\providecommand \href@noop [0]{\@secondoftwo}%
\providecommand \href [0]{\begingroup \@sanitize@url \@href}%
\providecommand \@href[1]{\@@startlink{#1}\@@href}%
\providecommand \@@href[1]{\endgroup#1\@@endlink}%
\providecommand \@sanitize@url [0]{\catcode `\\12\catcode `\$12\catcode
  `\&12\catcode `\#12\catcode `\^12\catcode `\_12\catcode `\%12\relax}%
\providecommand \@@startlink[1]{}%
\providecommand \@@endlink[0]{}%
\providecommand \url  [0]{\begingroup\@sanitize@url \@url }%
\providecommand \@url [1]{\endgroup\@href {#1}{\urlprefix }}%
\providecommand \urlprefix  [0]{URL }%
\providecommand \Eprint [0]{\href }%
\providecommand \doibase [0]{http://dx.doi.org/}%
\providecommand \selectlanguage [0]{\@gobble}%
\providecommand \bibinfo  [0]{\@secondoftwo}%
\providecommand \bibfield  [0]{\@secondoftwo}%
\providecommand \translation [1]{[#1]}%
\providecommand \BibitemOpen [0]{}%
\providecommand \bibitemStop [0]{}%
\providecommand \bibitemNoStop [0]{.\EOS\space}%
\providecommand \EOS [0]{\spacefactor3000\relax}%
\providecommand \BibitemShut  [1]{\csname bibitem#1\endcsname}%
\let\auto@bib@innerbib\@empty
\bibitem [{\citenamefont {'t~Hooft}(1974{\natexlab{a}})}]{tHooft:1974pnl}%
  \BibitemOpen
  \bibfield  {author} {\bibinfo {author} {\bibfnamefont {G.}~\bibnamefont
  {'t~Hooft}},\ }\href {\doibase 10.1016/0550-3213(74)90088-1} {\bibfield
  {journal} {\bibinfo  {journal} {Nucl. Phys. B}\ }\textbf {\bibinfo {volume}
  {75}},\ \bibinfo {pages} {461} (\bibinfo {year}
  {1974}{\natexlab{a}})}\BibitemShut {NoStop}%
\bibitem [{\citenamefont {Callan}\ \emph {et~al.}(1976)\citenamefont {Callan},
  \citenamefont {Coote},\ and\ \citenamefont {Gross}}]{Callan:1975ps}%
  \BibitemOpen
  \bibfield  {author} {\bibinfo {author} {\bibfnamefont {C.~G.}\ \bibnamefont
  {Callan}, \bibfnamefont {Jr.}}, \bibinfo {author} {\bibfnamefont
  {N.}~\bibnamefont {Coote}}, \ and\ \bibinfo {author} {\bibfnamefont {D.~J.}\
  \bibnamefont {Gross}},\ }\href {\doibase 10.1103/PhysRevD.13.1649} {\bibfield
   {journal} {\bibinfo  {journal} {Phys. Rev. D}\ }\textbf {\bibinfo {volume}
  {13}},\ \bibinfo {pages} {1649} (\bibinfo {year} {1976})}\BibitemShut
  {NoStop}%
\bibitem [{\citenamefont {Einhorn}(1976)}]{Einhorn:1976uz}%
  \BibitemOpen
  \bibfield  {author} {\bibinfo {author} {\bibfnamefont {M.~B.}\ \bibnamefont
  {Einhorn}},\ }\href {\doibase 10.1103/PhysRevD.14.3451} {\bibfield  {journal}
  {\bibinfo  {journal} {Phys. Rev. D}\ }\textbf {\bibinfo {volume} {14}},\
  \bibinfo {pages} {3451} (\bibinfo {year} {1976})}\BibitemShut {NoStop}%
\bibitem [{\citenamefont {Einhorn}\ \emph {et~al.}(1977)\citenamefont
  {Einhorn}, \citenamefont {Nussinov},\ and\ \citenamefont
  {Rabinovici}}]{Einhorn:1976ax}%
  \BibitemOpen
  \bibfield  {author} {\bibinfo {author} {\bibfnamefont {M.~B.}\ \bibnamefont
  {Einhorn}}, \bibinfo {author} {\bibfnamefont {S.}~\bibnamefont {Nussinov}}, \
  and\ \bibinfo {author} {\bibfnamefont {E.}~\bibnamefont {Rabinovici}},\
  }\href {\doibase 10.1103/PhysRevD.15.2282} {\bibfield  {journal} {\bibinfo
  {journal} {Phys. Rev. D}\ }\textbf {\bibinfo {volume} {15}},\ \bibinfo
  {pages} {2282} (\bibinfo {year} {1977})}\BibitemShut {NoStop}%
\bibitem [{\citenamefont {Brower}\ \emph {et~al.}(1977)\citenamefont {Brower},
  \citenamefont {Ellis}, \citenamefont {Schmidt},\ and\ \citenamefont
  {Weis}}]{Brower:1977hx}%
  \BibitemOpen
  \bibfield  {author} {\bibinfo {author} {\bibfnamefont {R.~C.}\ \bibnamefont
  {Brower}}, \bibinfo {author} {\bibfnamefont {J.~R.}\ \bibnamefont {Ellis}},
  \bibinfo {author} {\bibfnamefont {M.~G.}\ \bibnamefont {Schmidt}}, \ and\
  \bibinfo {author} {\bibfnamefont {J.~H.}\ \bibnamefont {Weis}},\ }\href
  {\doibase 10.1016/0550-3213(77)90303-0} {\bibfield  {journal} {\bibinfo
  {journal} {Nucl. Phys. B}\ }\textbf {\bibinfo {volume} {128}},\ \bibinfo
  {pages} {131} (\bibinfo {year} {1977})}\BibitemShut {NoStop}%
\bibitem [{\citenamefont {'t~Hooft}(1974{\natexlab{b}})}]{tHooft:1973alw}%
  \BibitemOpen
  \bibfield  {author} {\bibinfo {author} {\bibfnamefont {G.}~\bibnamefont
  {'t~Hooft}},\ }\href {\doibase 10.1016/0550-3213(74)90154-0} {\bibfield
  {journal} {\bibinfo  {journal} {Nucl. Phys. B}\ }\textbf {\bibinfo {volume}
  {72}},\ \bibinfo {pages} {461} (\bibinfo {year}
  {1974}{\natexlab{b}})}\BibitemShut {NoStop}%
\bibitem [{\citenamefont {Witten}(1979)}]{Witten:1979kh}%
  \BibitemOpen
  \bibfield  {author} {\bibinfo {author} {\bibfnamefont {E.}~\bibnamefont
  {Witten}},\ }\href {\doibase 10.1016/0550-3213(79)90232-3} {\bibfield
  {journal} {\bibinfo  {journal} {Nucl. Phys. B}\ }\textbf {\bibinfo {volume}
  {160}},\ \bibinfo {pages} {57} (\bibinfo {year} {1979})}\BibitemShut
  {NoStop}%
\bibitem [{\citenamefont {Bars}\ and\ \citenamefont
  {Green}(1978)}]{Bars:1977ud}%
  \BibitemOpen
  \bibfield  {author} {\bibinfo {author} {\bibfnamefont {I.}~\bibnamefont
  {Bars}}\ and\ \bibinfo {author} {\bibfnamefont {M.~B.}\ \bibnamefont
  {Green}},\ }\href {\doibase 10.1103/PhysRevD.17.537} {\bibfield  {journal}
  {\bibinfo  {journal} {Phys. Rev. D}\ }\textbf {\bibinfo {volume} {17}},\
  \bibinfo {pages} {537} (\bibinfo {year} {1978})}\BibitemShut {NoStop}%
\bibitem [{\citenamefont {Li}\ \emph {et~al.}(1987)\citenamefont {Li},
  \citenamefont {Wilets},\ and\ \citenamefont {Birse}}]{Li:1987hx}%
  \BibitemOpen
  \bibfield  {author} {\bibinfo {author} {\bibfnamefont {M.}~\bibnamefont
  {Li}}, \bibinfo {author} {\bibfnamefont {L.}~\bibnamefont {Wilets}}, \ and\
  \bibinfo {author} {\bibfnamefont {M.~C.}\ \bibnamefont {Birse}},\ }\href
  {\doibase 10.1088/0305-4616/13/7/005} {\bibfield  {journal} {\bibinfo
  {journal} {J. Phys. G}\ }\textbf {\bibinfo {volume} {13}},\ \bibinfo {pages}
  {915} (\bibinfo {year} {1987})}\BibitemShut {NoStop}%
\bibitem [{\citenamefont {Jia}\ \emph {et~al.}(2017)\citenamefont {Jia},
  \citenamefont {Liang}, \citenamefont {Li},\ and\ \citenamefont
  {Xiong}}]{Jia:2017uul}%
  \BibitemOpen
  \bibfield  {author} {\bibinfo {author} {\bibfnamefont {Y.}~\bibnamefont
  {Jia}}, \bibinfo {author} {\bibfnamefont {S.}~\bibnamefont {Liang}}, \bibinfo
  {author} {\bibfnamefont {L.}~\bibnamefont {Li}}, \ and\ \bibinfo {author}
  {\bibfnamefont {X.}~\bibnamefont {Xiong}},\ }\href {\doibase
  10.1007/JHEP11(2017)151} {\bibfield  {journal} {\bibinfo  {journal} {JHEP}\
  }\textbf {\bibinfo {volume} {11}},\ \bibinfo {pages} {151} (\bibinfo {year}
  {2017})},\ \Eprint {http://arxiv.org/abs/1708.09379} {arXiv:1708.09379
  [hep-ph]} \BibitemShut {NoStop}%
\bibitem [{\citenamefont {Hamer}(1982)}]{Hamer:1981yq}%
  \BibitemOpen
  \bibfield  {author} {\bibinfo {author} {\bibfnamefont {C.~J.}\ \bibnamefont
  {Hamer}},\ }\href {\doibase 10.1016/0550-3213(82)90009-8} {\bibfield
  {journal} {\bibinfo  {journal} {Nucl. Phys. B}\ }\textbf {\bibinfo {volume}
  {195}},\ \bibinfo {pages} {503} (\bibinfo {year} {1982})}\BibitemShut
  {NoStop}%
\bibitem [{\citenamefont {Berruto}\ \emph {et~al.}(2002)\citenamefont
  {Berruto}, \citenamefont {Giusti}, \citenamefont {Hoelbling},\ and\
  \citenamefont {Rebbi}}]{Berruto:2002gn}%
  \BibitemOpen
  \bibfield  {author} {\bibinfo {author} {\bibfnamefont {F.}~\bibnamefont
  {Berruto}}, \bibinfo {author} {\bibfnamefont {L.}~\bibnamefont {Giusti}},
  \bibinfo {author} {\bibfnamefont {C.}~\bibnamefont {Hoelbling}}, \ and\
  \bibinfo {author} {\bibfnamefont {C.}~\bibnamefont {Rebbi}},\ }\href
  {\doibase 10.1103/PhysRevD.65.094516} {\bibfield  {journal} {\bibinfo
  {journal} {Phys. Rev. D}\ }\textbf {\bibinfo {volume} {65}},\ \bibinfo
  {pages} {094516} (\bibinfo {year} {2002})},\ \Eprint
  {http://arxiv.org/abs/hep-lat/0201010} {arXiv:hep-lat/0201010} \BibitemShut
  {NoStop}%
\bibitem [{\citenamefont {Garc\'\i{}a~P\'erez}\ \emph
  {et~al.}(2016)\citenamefont {Garc\'\i{}a~P\'erez}, \citenamefont
  {Gonz\'alez-Arroyo}, \citenamefont {Keegan},\ and\ \citenamefont
  {Okawa}}]{GarciaPerez:2016hph}%
  \BibitemOpen
  \bibfield  {author} {\bibinfo {author} {\bibfnamefont {M.}~\bibnamefont
  {Garc\'\i{}a~P\'erez}}, \bibinfo {author} {\bibfnamefont {A.}~\bibnamefont
  {Gonz\'alez-Arroyo}}, \bibinfo {author} {\bibfnamefont {L.}~\bibnamefont
  {Keegan}}, \ and\ \bibinfo {author} {\bibfnamefont {M.}~\bibnamefont
  {Okawa}},\ }\href {\doibase 10.22323/1.256.0337} {\bibfield  {journal}
  {\bibinfo  {journal} {PoS}\ }\textbf {\bibinfo {volume} {LATTICE2016}},\
  \bibinfo {pages} {337} (\bibinfo {year} {2016})},\ \Eprint
  {http://arxiv.org/abs/1612.07380} {arXiv:1612.07380 [hep-lat]} \BibitemShut
  {NoStop}%
\bibitem [{\citenamefont {Willemsen}(1978)}]{Willemsen:1977fr}%
  \BibitemOpen
  \bibfield  {author} {\bibinfo {author} {\bibfnamefont {J.~F.}\ \bibnamefont
  {Willemsen}},\ }\href {\doibase 10.1103/PhysRevD.17.574} {\bibfield
  {journal} {\bibinfo  {journal} {Phys. Rev.}\ }\textbf {\bibinfo {volume}
  {D17}},\ \bibinfo {pages} {574} (\bibinfo {year} {1978})}\BibitemShut
  {NoStop}%
\bibitem [{\citenamefont {Bjorken}(1982)}]{Bjorken:1979hv}%
  \BibitemOpen
  \bibfield  {author} {\bibinfo {author} {\bibfnamefont {J.~D.}\ \bibnamefont
  {Bjorken}},\ }\enquote {\bibinfo {title} {Elements of quantum
  chromodynamics},}\ in\ \href {\doibase 10.1007/978-1-4899-6691-9_5} {\emph
  {\bibinfo {booktitle} {Lectures on Lepton Nucleon Scattering and Quantum
  Chromodynamics}}}\ (\bibinfo  {publisher} {Birkh{\"a}user Boston},\ \bibinfo
  {address} {Boston, MA},\ \bibinfo {year} {1982})\ pp.\ \bibinfo {pages}
  {423--561}\BibitemShut {NoStop}%
\bibitem [{\citenamefont {Leibbrandt}(1987)}]{Leibbrandt:1987qv}%
  \BibitemOpen
  \bibfield  {author} {\bibinfo {author} {\bibfnamefont {G.}~\bibnamefont
  {Leibbrandt}},\ }\href {\doibase 10.1103/RevModPhys.59.1067} {\bibfield
  {journal} {\bibinfo  {journal} {Rev. Mod. Phys.}\ }\textbf {\bibinfo {volume}
  {59}},\ \bibinfo {pages} {1067} (\bibinfo {year} {1987})}\BibitemShut
  {NoStop}%
\bibitem [{\citenamefont {Strocchi}(2013)}]{Strocchi:2013awa}%
  \BibitemOpen
  \bibfield  {author} {\bibinfo {author} {\bibfnamefont {F.}~\bibnamefont
  {Strocchi}},\ }\href
  {https://doi.org/10.1093/acprof:oso/9780199671571.001.0001} {\emph {\bibinfo
  {title} {{An introduction to non-perturbative foundations of quantum field
  theory}}}}\ (\bibinfo  {publisher} {Oxford University Press},\ \bibinfo
  {year} {2013})\BibitemShut {NoStop}%
\bibitem [{\citenamefont {Christ}\ and\ \citenamefont
  {Lee}(1980)}]{Christ:1980ku}%
  \BibitemOpen
  \bibfield  {author} {\bibinfo {author} {\bibfnamefont {N.~H.}\ \bibnamefont
  {Christ}}\ and\ \bibinfo {author} {\bibfnamefont {T.~D.}\ \bibnamefont
  {Lee}},\ }\href {\doibase 10.1103/PhysRevD.22.939} {\bibfield  {journal}
  {\bibinfo  {journal} {Phys. Rev.}\ }\textbf {\bibinfo {volume} {D22}},\
  \bibinfo {pages} {939} (\bibinfo {year} {1980})}\BibitemShut {NoStop}%
\bibitem [{\citenamefont {Weinberg}(2005)}]{Weinberg:1995mt}%
  \BibitemOpen
  \bibfield  {author} {\bibinfo {author} {\bibfnamefont {S.}~\bibnamefont
  {Weinberg}},\ }\href@noop {} {\emph {\bibinfo {title} {{The Quantum theory of
  fields. Vol. 1: Foundations}}}}\ (\bibinfo  {publisher} {Cambridge University
  Press},\ \bibinfo {year} {2005})\BibitemShut {NoStop}%
\bibitem [{\citenamefont {Salpeter}\ and\ \citenamefont
  {Bethe}(1951)}]{Salpeter:1951sz}%
  \BibitemOpen
  \bibfield  {author} {\bibinfo {author} {\bibfnamefont {E.~E.}\ \bibnamefont
  {Salpeter}}\ and\ \bibinfo {author} {\bibfnamefont {H.~A.}\ \bibnamefont
  {Bethe}},\ }\href {\doibase 10.1103/PhysRev.84.1232} {\bibfield  {journal}
  {\bibinfo  {journal} {Phys. Rev.}\ }\textbf {\bibinfo {volume} {84}},\
  \bibinfo {pages} {1232} (\bibinfo {year} {1951})}\BibitemShut {NoStop}%
\bibitem [{\citenamefont {Chodos}\ \emph {et~al.}(1974)\citenamefont {Chodos},
  \citenamefont {Jaffe}, \citenamefont {Johnson}, \citenamefont {Thorn},\ and\
  \citenamefont {Weisskopf}}]{Chodos:1974je}%
  \BibitemOpen
  \bibfield  {author} {\bibinfo {author} {\bibfnamefont {A.}~\bibnamefont
  {Chodos}}, \bibinfo {author} {\bibfnamefont {R.}~\bibnamefont {Jaffe}},
  \bibinfo {author} {\bibfnamefont {K.}~\bibnamefont {Johnson}}, \bibinfo
  {author} {\bibfnamefont {C.~B.}\ \bibnamefont {Thorn}}, \ and\ \bibinfo
  {author} {\bibfnamefont {V.}~\bibnamefont {Weisskopf}},\ }\href {\doibase
  10.1103/PhysRevD.9.3471} {\bibfield  {journal} {\bibinfo  {journal} {Phys.
  Rev. D}\ }\textbf {\bibinfo {volume} {9}},\ \bibinfo {pages} {3471} (\bibinfo
  {year} {1974})}\BibitemShut {NoStop}%
\bibitem [{\citenamefont {Hoyer}(2021)}]{Hoyer:2021adf}%
  \BibitemOpen
  \bibfield  {author} {\bibinfo {author} {\bibfnamefont {P.}~\bibnamefont
  {Hoyer}},\ }\href {\doibase 10.1007/978-3-030-79489-7} {\emph {\bibinfo
  {title} {{Journey to the Bound States}}}},\ SpringerBriefs in Physics\
  (\bibinfo  {publisher} {Springer},\ \bibinfo {year} {2021})\ \Eprint
  {http://arxiv.org/abs/2101.06721} {arXiv:2101.06721 [hep-ph]} \BibitemShut
  {NoStop}%
\bibitem [{\citenamefont {Peskin}\ and\ \citenamefont
  {Schroeder}(1995)}]{Peskin:1995ev}%
  \BibitemOpen
  \bibfield  {author} {\bibinfo {author} {\bibfnamefont {M.~E.}\ \bibnamefont
  {Peskin}}\ and\ \bibinfo {author} {\bibfnamefont {D.~V.}\ \bibnamefont
  {Schroeder}},\ }\href
  {https://www.taylorfrancis.com/books/mono/10.1201/9780429503559/introduction-quantum-field-theory-michael-peskin}
  {\emph {\bibinfo {title} {{An Introduction to quantum field theory}}}}\
  (\bibinfo  {publisher} {Addison-Wesley},\ \bibinfo {address} {Reading, USA},\
  \bibinfo {year} {1995})\BibitemShut {NoStop}%
\bibitem [{\citenamefont {Dietrich}\ \emph {et~al.}(2013)\citenamefont
  {Dietrich}, \citenamefont {Hoyer},\ and\ \citenamefont
  {J{\"a}rvinen}}]{Dietrich:2012un}%
  \BibitemOpen
  \bibfield  {author} {\bibinfo {author} {\bibfnamefont {D.~D.}\ \bibnamefont
  {Dietrich}}, \bibinfo {author} {\bibfnamefont {P.}~\bibnamefont {Hoyer}}, \
  and\ \bibinfo {author} {\bibfnamefont {M.}~\bibnamefont {J{\"a}rvinen}},\
  }\href {\doibase 10.1103/PhysRevD.87.065021} {\bibfield  {journal} {\bibinfo
  {journal} {Phys. Rev.}\ }\textbf {\bibinfo {volume} {D87}},\ \bibinfo {pages}
  {065021} (\bibinfo {year} {2013})},\ \Eprint {http://arxiv.org/abs/1212.4747}
  {arXiv:1212.4747 [hep-ph]} \BibitemShut {NoStop}%
\bibitem [{\citenamefont {Dietrich}\ \emph {et~al.}(2012)\citenamefont
  {Dietrich}, \citenamefont {Hoyer},\ and\ \citenamefont
  {J{\"a}rvinen}}]{Dietrich:2012iy}%
  \BibitemOpen
  \bibfield  {author} {\bibinfo {author} {\bibfnamefont {D.~D.}\ \bibnamefont
  {Dietrich}}, \bibinfo {author} {\bibfnamefont {P.}~\bibnamefont {Hoyer}}, \
  and\ \bibinfo {author} {\bibfnamefont {M.}~\bibnamefont {J{\"a}rvinen}},\
  }\href {\doibase 10.1103/PhysRevD.85.105016} {\bibfield  {journal} {\bibinfo
  {journal} {Phys. Rev.}\ }\textbf {\bibinfo {volume} {D85}},\ \bibinfo {pages}
  {105016} (\bibinfo {year} {2012})},\ \Eprint {http://arxiv.org/abs/1202.0826}
  {arXiv:1202.0826 [hep-ph]} \BibitemShut {NoStop}%
\bibitem [{\citenamefont {Plesset}(1932)}]{Plesset:1930zz}%
  \BibitemOpen
  \bibfield  {author} {\bibinfo {author} {\bibfnamefont {M.~S.}\ \bibnamefont
  {Plesset}},\ }\href {\doibase 10.1103/PhysRev.41.278} {\bibfield  {journal}
  {\bibinfo  {journal} {Phys. Rev.}\ }\textbf {\bibinfo {volume} {41}},\
  \bibinfo {pages} {278} (\bibinfo {year} {1932})}\BibitemShut {NoStop}%
\bibitem [{\citenamefont {Melnitchouk}\ \emph {et~al.}(2005)\citenamefont
  {Melnitchouk}, \citenamefont {Ent},\ and\ \citenamefont
  {Keppel}}]{Melnitchouk:2005zr}%
  \BibitemOpen
  \bibfield  {author} {\bibinfo {author} {\bibfnamefont {W.}~\bibnamefont
  {Melnitchouk}}, \bibinfo {author} {\bibfnamefont {R.}~\bibnamefont {Ent}}, \
  and\ \bibinfo {author} {\bibfnamefont {C.}~\bibnamefont {Keppel}},\ }\href
  {\doibase 10.1016/j.physrep.2004.10.004} {\bibfield  {journal} {\bibinfo
  {journal} {Phys. Rept.}\ }\textbf {\bibinfo {volume} {406}},\ \bibinfo
  {pages} {127} (\bibinfo {year} {2005})},\ \Eprint
  {http://arxiv.org/abs/hep-ph/0501217} {arXiv:hep-ph/0501217} \BibitemShut
  {NoStop}%
\bibitem [{\citenamefont {Kronfeld}(2012)}]{Kronfeld:2012uk}%
  \BibitemOpen
  \bibfield  {author} {\bibinfo {author} {\bibfnamefont {A.~S.}\ \bibnamefont
  {Kronfeld}},\ }\href {\doibase 10.1146/annurev-nucl-102711-094942} {\bibfield
   {journal} {\bibinfo  {journal} {Ann. Rev. Nucl. Part. Sci.}\ }\textbf
  {\bibinfo {volume} {62}},\ \bibinfo {pages} {265} (\bibinfo {year} {2012})},\
  \Eprint {http://arxiv.org/abs/1203.1204} {arXiv:1203.1204 [hep-lat]}
  \BibitemShut {NoStop}%
\bibitem [{\citenamefont {Hoyer}(2023)}]{Hoyer:2023noz}%
  \BibitemOpen
  \bibfield  {author} {\bibinfo {author} {\bibfnamefont {P.}~\bibnamefont
  {Hoyer}},\ }\href {\doibase 10.1103/PhysRevD.108.034031} {\bibfield
  {journal} {\bibinfo  {journal} {Phys. Rev. D}\ }\textbf {\bibinfo {volume}
  {108}},\ \bibinfo {pages} {034031} (\bibinfo {year} {2023})},\ \Eprint
  {http://arxiv.org/abs/2304.11903} {arXiv:2304.11903 [hep-ph]} \BibitemShut
  {NoStop}%
\end{thebibliography}%
\end{document}